\documentclass[aps,a4paper,floatfix,nofootinbib,unsortedaddress]{revtex4-2}
\usepackage{packages}
\usepackage{mycommands}
\begin{document}
\title{
Discrete-to-Continuum Approach for the Analytic Continuation of One-Particle Propagator on the Circle  
}
\author{Andrea Stampiggi}\email{astampig@sissa.it} 
\affiliation{SISSA and INFN, Sezione di Trieste, via Bonomea 265, I-34136, Trieste, Italy}
\date{\today}
\begin{abstract}
Despite the simplicity of one-particle dynamics, explicit expressions for the one-dimensional propagator on a circle suitable to numerical evaluation are surprisingly lacking -- not only in the presence of potentials but even in the free case. 
Using a lattice regularization of the circle, we derive finite expressions for the free discrete propagator through an algebraic approach, aiming to provide physical insight into the readout of a digital quantum simulation. Moreover, these expressions allow for the reconstruction of the propagator in the continuous circle limit, which exhibits in the free case a peculiar non-analytic behavior in its transition between irrational and rational times. The latter propagator yields a finite analytic continuation of the corresponding elliptic theta function at the locus of essential singularities for real times, achieved through the introduction of a $\sigma$ distribution -- the ``square-root'' of the Dirac delta. We also show that the well-known infinite line limit is consistently recovered within this approach. In addition, we apply these results by studying numerically the dynamics of wave packets in cosine and random potentials. At early simulation times, we observe evidence of the semi-classical limit, where the probability density maximum follows the minimum of the propagator phase.  

\end{abstract}
\maketitle
\section{Introduction}\label{s_intro}
It is well known that the one-dimensional propagator of a quantum particle on an infinite continuous line, transitioning from position $x$ to $x'$ in time $t$, assumes a surprisingly simple expression in terms of a path integral \cite{feynman1948path, dirac1945, feynman2010quantum} in which the contributions are from all the classical trajectories $\chi(\tau)$ such that $\chi(0)=x$ and $\chi(t) = x'$ with classical action $S_{x}^{x'}(\chi(\tau), t)$:
\begin{equation}\label{eq_propagator_continuum_intro}
    \braket{x'| e^{-i H t/\hbar} | x} = \inv{Z_t} \int_{\chi(0) = x}^{\chi(t) = x'} \dd \chi(\tau) e^{i S_{x}^{x'}(\chi(\tau), t)/\hbar}.
\end{equation}
Here $H$ is the one-body Hamiltonian and $Z_t$ is a constant fixed by requiring that the propagator is the generator of time evolution on any physical wave function. Also, in the semi-classical limit of vanishing Planck’s constant $\hbar$, the most probable position to occur in a measurement (the maximum of probability density) will follow the trajectory dictated by the minimum of the propagator phase. 

From the analytical perspective, the computation of the one-particle propagator phase in the infinite line requires the integration of a Lagrangian, which is often the starting point in quantum field theory. In many-body physics, where interactions between different constituents generate nontrivial dynamics, the Hamiltonian approach seems the most natural one. Lagrangian and Hamiltonian formalisms are equivalent, and it is often possible to obtain the Hamiltonian from the Lagrangian when the system evolves on an infinite continuum. However, the latter approach can not be applied in the presence of constraints \cite{Dirac_1950,Dirac_1951}, the simplest example being the restriction of the dynamics to a compact geometry, where physical momenta become discrete. For the simplest one-dimensional case, i.e. a circle of length $L$, the one-particle free propagator is 
\begin{equation}\label{e_propa_circle_formal}
\mathcal{D}_{x}^{x'} (t) = \sum_{\kappa \in \momentumlattice{L}} \frac{e^{i \kappa (x'-x)}}{L}\exp\br{ - i \frac{\hbar}{2m} \kappa^2 t}.
\end{equation}
For generic displacements and real physical times, the sum does not converge due to the interference of infinitely many waves.

Beyond numerical solutions of the Schr\"{o}dinger equation, constructed from the Hamiltonian in presence of constraints \cite{SCARDICCHIO20027}, the knowledge of the propagator is essential in providing alternative approaches to the study of dynamics. In the literature, the one-dimensional particle on the circle has been studied starting from the path integral formalism on the infinite line, both at imaginary \cite{simanek} and real \cite{schulman2012techniques} times. Both approaches start from different considerations and do not provide propagators which are suitable to numerical evaluations. Regarding the first case \cite{simanek}, the analytic continuation from imaginary to real times of the propagator is allowed in the line but not on the circle. In fact, on the line the propagator assumes contributions from a continuum of momenta, while one the circle the infinite sum of phases in Eq.~\eqref{e_propa_circle_formal} is an elliptic function evaluated at the locus of essential singularities. As such, no straightforward analytic continuation from imaginary to real times can be performed. On the other hand, the propagator of \cite{schulman2012techniques} explicitly features the presence of the elliptic theta function at real times and a phase, acquired by the wave function upon periodic shifts. While no procedure is given to fix this phase, even in the simplest case of a free particle, it is noticed that the free propagator exhibits nontrivial zeros depending on whether time is rational or irrational, without discussing the implications on the dynamics. Using a complementary approach, this work recovers and provides a precise characterization of this intriguing non-analytic behavior in time, which is a feature of the circle and is absent in the line.

The issue of obtaining and analyzing finite expressions for the one-dimensional one-particle propagator, either free or in presence of a nonzero potential, also finds relevance with the advent of new quantum technologies, in particular quantum computation. Recently, an algorithm of digital quantum simulation for measuring reflection and transmission amplitude for the scattering with a short-ranged potential through real-time evolution has been proposed \cite{MST2024}. Even though the latter algorithm explicitly tackles the problem of simulating particle dynamics by encoding the physical degrees of freedom into quantum registers, information about dynamics can be also obtained by solving the Schr\"{o}dinger equation as an ordinary linear differential equation. Relying on the matrix inversion technique, attempts in this direction have been made \cite{LloydLinear, SommaAdiabatic, LloydDifferentialInhomogeneous}. The inversion process naturally involves the presence of an inhomogeneous term, and finds application in solving the Schr\"{o}dinger equation in presence of source terms \cite{NumericalSchrodingerSource}. However, such algorithms often lack in physical intuition, and analytical tools --such as finite expressions for the particle propagator -- can provide valuable insights into the collective evolution of the simulating platform, i.e., the many-body qubit system.

In simulating one-dimensional dynamics, physical space is naturally cut off to some length scale $L$, which sets the length of a circle, with the particle wave function being periodic over it. In the setting of digital quantum simulations it is natural to divide it into $N_n = 2^n$ intervals of length $\ell = L/ 2^n $, where $n$ is the number of qubits in the register. One of the main results of this paper is the exact form of the free propagator for finite $N_n$ and $L$ for certain values of the simulation time $\gt = \frac{\hbar}{m} \frac{2\pi}{L^2} t$, $m$ being the mass of the particle and $t$ the physical time. We show that for the special value $\gt_n = 2^{-n}$, and multiples thereof, the outcomes of the propagator at finite $N_n$ coincide with those in the continuum ($N_n\to \infty$). As such, $\gt_n$ is the smallest natural time step, given $n$. While the free propagator is only given at a specific time, it is sufficient to numerically simulate one-dimensional Hamiltonians composed of a free and potential term for all simulation times which are integer multiples of $\gt_n$, giving access to the study of finite-size effects characteristic of compact geometries. 

It is well known that, on the line, free propagation of a wave packet implies the spreading of the wave function and therefore the delocalization in position space of the corresponding particle. On the other hand, on the circle, free dynamics is periodic in time. Therefore a free periodic packet first delocalizes at early simulation times, but as time approaches the period, dynamics is enriched by a competition effect, since the packet will ``relocalize'' to its zero-time wave function. This nontrivial dynamics is not only a feature of early and late times, but it is also possible to observe a rich structure of the periodic wave function at intermediate times, quantified by the number of probability maxima, indicating that the particle has regions in the circle where it is more probable to be found. This behavior is paradigmatic in the case of a free ideal particle, for which we provide a more quantitative analysis of this phenomenon.

Since the simulation time $\gt$ is proportional to the Plank's constant, it is possible to observe the evidence of the semi-classical limit. Here the the probability maximum dictates the trajectory of the semi-classical trajectory and follows the minimum of the propagator's phase. The knowledge of the free propagator alone allows for the numerical check of this behavior for free wave packets highly peaked in momentum space. It is also possible to introduce nonzero potentials and verify numerically that the region of early simulation times indeed presents the evidence of the semi-classical limit, by comparing the maximum of probability and the wave function phase.

We also show how the free propagator on the discretized circle can be analytically continued to the continuum. The discrete-to-continuum limit ($N\to\infty$) requires the mapping of several objects, such as position and momenta states and their eigenvalues, from a periodic finite lattice of spacing $\ell = L/N$ to the continuum. Generally, the free propagator on a circle is given in terms of a distribution $\sigma$, which can be seen as the ``square root'' of the Dirac $\delta$. Starting from a localized state on the circle, when the simulation time is a rational number, $\gt = p/q$, $p$ and $q$ coprimes, at most $q$ points are found as outcomes of the measurement process. This is in contrast to the expected behavior on the line where, no matter the time $t$, even if infinitesimally small, free propagation of a localized state will yield a uniform probability of finding a particle in some other position. This is actually what happens on the circle (finite $L$) when $\gt$ is irrational. As far as the phase of the propagator is concerned, at $\gt = 1/q$ it is analogous to that of a particle on the line. For $\gt=p/q$ the phase seems to be more complicated and no simple expressions were found. For $\gt$ irrational, the phase is not an analytic continuous function of displacement. The only case which is analytically tractable is the limit of infinitesimal $\gt$, which leads consistently to the the free propagator on the line. Indeed, since the simulation time $\gt$ is inversely proportional to $L^2$, infinitesimal $\gt$'s correspond to the limit $L\to\infty$, i.e. the line, for any nonzero time $t$. As such, the propagator on the circle contains information about that on the line, but it is not possible to recover the latter from the former. In fact the order of the limits ($N$ and $L\to\infty$) can not be reversed for this problem: first from the finite lattice to a finite continuum ($N\to\infty$) and then from the finite continuum to the infinite continuum ($L\to\infty$).

The content of this paper is organized as follows. In Section~\ref{s_rev_oneparticle} we review one-particle dynamics and the problem of time evolution. Section~\ref{s_rev_algo} reviews the digital quantum simulation of one-dimensional one-particle dynamics with a system of qubits and remarks how the discrete one-particle propagator emerges naturally in this setting. In Section~\ref{s_propa_circle_finite_N} we provide the derivation of the free propagator on the circle with a finite amount of sampled points $N$. In Section~\ref{s_propa_circle_continuum} we address the limits $N\to \infty$ and $L\to\infty$. Lastly, Section~\ref{s_numerical} is dedicated to the numerical investigation of the time-evolution for different potentials. Conclusive remarks are gathered in Section~\ref{s_conclusions}.

The paper also presents several appendices, aimed at complementing the discussion of the main text.
In Appendix~\ref{a_uncertainty} we give a self-contained presentation of the one-particle Hilbert space on several geometries: line, segment, circle and lattice. Most importantly, we show that momenta defined on these geometries indeed satisfy the same uncertainty relations one would normally expect when acting on physical states. Appendix~\ref{a_technical} is dedicated to a technical result which has been left out from Section~\ref{s_propa_circle_finite_N}. In Appendix~\ref{a_adimensional_free_hamiltonian} we propose a free Hamiltonian whose momenta are adimensional and spaced with the inverse of the number of qubit states and study its propagator. Even though this Hamiltonian does not have at the moment a clear physical interpretation, its propagator features interesting properties.

\section{Review of One-Particle Dynamics}\label{s_rev_oneparticle}

In this section we provide a short review of one-particle dynamics on the continuum. The dynamics in a discrete setting, natural in the context of numerical simulation, will be considered in the next.

The system we would like to simulate is a physical particle of mass $m$ interacting with a one-dimensional potential $V(x)$. In absence of potential or in presence of potentials of the type $\sum_{i=1}^N V_i(x_i)$, generalizations to higher dimensions is straightforward. Since in any computation, the whole line can not be sampled, but only extrapolated, we restrict the simulation to a finite length $L$. If a wave function to be simulated is provided, $L$ can be chosen such that the truncation to a finite volume is negligible \cite{MST2024}. For the present purposes, we let $L$ be arbitrary and not tied to the specific state to be simulated. 

A one-particle system in an initial state $\ket{\Psi}$ will evolve according to an Hamiltonian $H_L$ as
\begin{equation}\label{e_evo_state_circle}
    \ket{\Psi_t} = e^{- \frac{i}{\hbar} H_L t} \ket{\Psi}.
\end{equation}
Introducing physical bases, let $\ket{x}$ be an eigenstate of position and $\ket{\kappa}$ be an eigenstate of momentum. On a circle of length $L$, the outcome of a position measurement will yield $x \in [-L/2, L/2)$ -- or any other interval which is obtained through a shift $x_0 \in [0, L)$. Therefore the position operator $\hat{X}$ has continuous spectrum. On the other hand, momenta are discrete and spaced by a quantity $2\pi/L$. The eigenvalues are 
\begin{equation}
    \hat{\kappa} \ket{\kappa_M} = \kappa_M \ket{\kappa_M}, \quad \kappa_M = \frac{2\pi}{L} M, \quad M\in \integers.
\end{equation}
Crucially, it is only on the Hilbert space of periodic wave functions ($\mathcal{H}_{L}^{\text{P}}$)
\begin{equation}
    \braket{x|\Psi} = \psi(x) = \psi(x + nL), \quad n\in\integers,
\end{equation}
that the basis $\ket{\kappa_M}$ is conjugate to that of position $\ket{x}$, so that $\kappa_M$ can be interpreted as momentum, see Appendix~\ref{a_uncertainty} for details.

As a consequence, the physically meaningful potentials for the time evolution must also be periodic. One important example we will consider in the following is 
\begin{equation}\label{e_harmonic_potential}
    V^{\text{H.O.}}_L (x) = m \omega^2 L^2 \br{1-\cos\br{\frac{2\pi}{L} x}},
\end{equation}
which is of period $L$ in position and yields the quadratic potential in the limit $L\to\infty$. Here $m$ is the mass of the particle and $\omega$ is a constant of dimension of inverse time $\ut^{-1}$.

As far as the kinetic term is concerned, it is diagonal in the basis of momenta and the eigenvalues are proportional to $\kappa_M^2$. Written as operator, it reads
\begin{equation}
    K_L = \sum_{M\in \integers} \frac{\hbar^2}{2m}\br{\frac{2\pi}{L}}^2 M^2 \pro{M} \coinc \sum_{\kappa \in \momentumlattice{L}} \frac{\hbar^2}{2m}\kappa^2 \pro{\kappa}.
\end{equation}

In order to pass from momentum to position basis and vice versa, we introduce a unitary mapping between the two, the "quantum Fourier Transform" (qFT) $\hat{\mathcal{F}}$, whose matrix elements are
\begin{equation}
    \hat{\mathcal{F}}_x^\kappa = \braket{\kappa|x} = \inv{\sqrt{L}} e^{-i \kappa x}, \quad \kappa \in \momentumlattice{L}, \quad x \in [-L/2, L/2),
\end{equation}
which is also present in the theory of quantum computation \cite{r_2010nielsen_b_qc_qi}, as we will argue in the next section. Since the matrix is unitary, $\hat{\mathcal{F}}^{-1} = \hat{\mathcal{F}}^\dagger$, the matrix elements of the inverse read 
\begin{equation}
    \br{\hat{\mathcal{F}}^\dagger}^x_\kappa = \braket{x|\kappa} = \inv{\sqrt{L}} e^{i \kappa x}, \quad \kappa \in \momentumlattice{L}, \quad x \in [-L/2, L/2).
\end{equation}
In other words, we can write any eigenstate of position as uniform superposition of eigenstates of momentum and vice-versa:
\begin{equation}\label{e_qFT_circle}
    \ket{x} = \sum_{\kappa \in \momentumlattice{L}} \inv{\sqrt{L}} e^{- i \kappa x} \ket{\kappa}, \quad \ket{\kappa} = \int_{-L/2}^{L/2} \dd x\, \inv{\sqrt{L}} e^{i\kappa x} \ket{x}.
\end{equation}
Given a physical state $\ket{\Psi}$ on $\mathcal{H}_{L}^{\text{P}}$, the qFT acts on the wave functions as the ordinary Fourier Transform:
\begin{equation}
    \braket{x|\Psi} = \psi(x) = \sum_{\kappa\in \momentumlattice{L}} \inv{\sqrt{L}} \Psi_\kappa e^{i\kappa x}, \quad \braket{\kappa|\Psi} = \Psi_\kappa = \int_{-L/2}^{L/2} \dd x\, \inv{\sqrt{L}} \psi(x) e^{-i \kappa x}.
\end{equation}
From the latter relations it is evident that the position wave function must be periodic in $L$.

Let us turn back to Eq.~\eqref{e_evo_state_circle}. We would like to study the position wave function after the evolution of time $t$:
\begin{equation}\label{e_time_evo_wave_f_circle}
    \psi_t(x') = \int_L \dd x\, \braket{x'| e^{- \frac{i}{\hbar} H_L t}| x} \psi(x) \coinc \int_L \dd x\, \br{D_{H_L}(t)}_x^{x'} \psi(x),
\end{equation}
where $\int_L$ denotes an integration over one period of the circle. Here $\br{D_{H_L}(t)}_x^{x'} $ is the propagator associated to the Hamiltonian $H_L$, which is made up of kinetic $K_L$ and potential $V$ operators which are diagonal in the basis of momenta and position, respectively. In addressing the time evolution, in light of the linearity of the Schr\"{o}dinger equation, the propagator completely solves the problem. It also has an important physical interpretation: it is the probability amplitude that a plane wave $\ket{x}$ is found at time $t$ at $\ket{x'}$ if it is evolved with $H_L$. 

Since $L$ is finite, we are certain that after a time $t$ the particle is always found somewhere on the circle, for it can not escape. Naively, one would be led to assume conservation of total probability, as a consequence. However, as it is the case in the well-known continuum line, the overall probability is a divergent quantity. It is only the probability density  $\norm{\br{D_{H_L}(t)}_x^{x'}}$ which is finite for a continuum of states.
\comment{
\begin{equation}
    1 \coinc \int_{-L/2}^{L/2} \dd (x'-x) \norm{\br{D_{H_L}(t)}_x^{x'}}, \quad\forall t.
\end{equation}
This property is no longer true in the line limit $L\to\infty$.} As evidenced in the later discussion, for some times the spectrum of the position operator becomes discrete and it is legitimate to require conservation of total probability. Henceforth, when simulating the dynamics of a particle on a continuum (line or circle) partitioned into a finite amount of intervals, conservation of probability on the corresponding propagator is a property to be required. In the following we will omit the subscript referring to $H_L$ in the propagator, denoting it simply by $D$.

\begin{figure}
    \centering
    \includegraphics[scale =0.66]{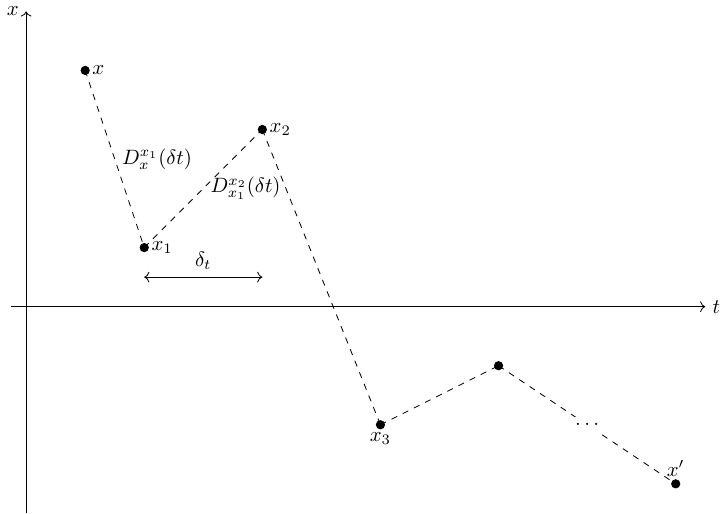}
    \caption{Schematic representation of a path contributing to Eq.~\eqref{e_path_integral_circle_propa}, connecting the initial and finite points, $x$ and $x'$ respectively, in a finite time $t$. A trajectory is obtained by partitioning the interval in smaller pieces of length $\delta t$ and is labeled by the sequence $\cbr{x, x_1, x_2, x_3, \ldots, x'}$.   
    In general, each trajectory contributes to the overall probability density amplitude $D_{x}^{x'}(t)$, generating nontrivial interference patterns characteristic of quantum particle dynamics.}
    \label{f_path_integral}
\end{figure}

Stemming from the latter considerations, we can derive another important property. Suppose that in the time evolution, we consider an intermediate time $t' \leq t$. As pointed out by \cite{feynman1948path}, if we make a measurement at $t'$, we would inevitably destroy interference effects. It is then important not to affect the particle in any way at the intermediate time. We can then imagine that at time $t'$ the particle can propagate to all points on the circle with wave function $\psi_{t'}(x'')$, given also by Eq.~\eqref{e_time_evo_wave_f_circle}. As such, the propagator must satisfy, for any intermediate time,
\begin{equation}\label{e_convo_propagator}
    D_{x}^{x'}(t) = \int_{L} \dd x'' \, D_{x''}^{x'}(t-t') D_{x}^{x''}(t'), \quad t'\leq t.
\end{equation}
If the interval $t$ is divided into $N_t$ steps of width $\delta t$, then we can generalize the last equation to 
\begin{equation}\label{e_path_integral_circle_propa}
    D_{x}^{x'}(t) = \int_L \dd x_1 \cdots \int_L \dd x_{N_t}\,  D_{x_{N_t}}^{x'}(\delta t) D_{x_{N_t -1}}^{x_{N_t}}(\delta t)\ldots D_{x_1}^{x_2}(\delta t)  D_{x_1}^{x_2}(\delta t) D_{x}^{x_1}(\delta t),
\end{equation}
pictorially visualized in Fig.~\ref{f_path_integral}. As such, the propagation of a single particle can be seen as a sum over all possible paths allowed at intermediate times. It contains within its bosom all the relevant theory of path integrals for non-relativistic quantum mechanics. It has been shown that the path integral formalism leads essentially to the same quantum mechanics in the pictures of Schr\"{o}dinger and Heisenberg \cite{feynman1948path}.

The latter equation has been presented in a form which is appealing when the circle is not continuous, but is discretized into a finite set of equally spaced points: integrals become sums and we can see path integration as a sequence of matrix products. This observation will be crucial in bridging Sec.~\ref{s_propa_circle_finite_N} to Sec.~\ref{s_propa_circle_continuum}.

Moreover, if the time is sufficiently small, $\delta t \ll t$, we can treat the infinitesimal time evolution operator generated by $H_L$ as the product of two commuting unitaries, generated by the kinetic and potential operators:
\begin{equation} \label{e_propa_infinitesimal_circle}
    D_{x}^{x'}(\delta t) = \int_L \dd x'' \braket{x'|e^{-\frac{i}{\hbar} V \delta t}|x''}\braket{x''| e^{-\frac{i}{\hbar} K_L \delta t}|x} + O(\delta t) \coinc \int_L \dd x'' \br{e^{-\frac{i}{\hbar} V\delta t}}^{x'}_{x''} \mathcal{D}_{x}^{x''} (\delta t) + O(\delta t).
\end{equation}
Here $\mathcal{D}_{x}^{x'} (\delta t)$ is the free physical propagator on the circle. On the line, we know that this propagator can be given meaning in the sense of analytic continuation of the Gaussian integral
\begin{equation}\label{e_propa_line_limit}
    \Lim{L}{\infty} \mathcal{D}_{x}^{x'} (\delta t) = \sqrt{\frac{m}{i 2\pi \hbar \delta t} }\exp\br{\frac{i}{\hbar} m \frac{(x'-x)^2}{2 \delta t}},
\end{equation}
while on a circle, it is given by Eq.~\eqref{e_propa_circle_formal}, due to momenta discretization. For the latter case, for generic displacements and real times the summation will in general lead to divergent quantities, and there is no hope that a naive truncation can lead to definite results, since we are summing phases and no momenta periodicity is expected for arbitrary $t$. Another way of seeing this is noticing that the free propagator is related to the elliptic theta function
\begin{equation}\label{e_adim_time}
    \theta(\xi; \gt) = \sum_{M\in\integers} \exp\br{-i 2\pi M^2 \frac{\gt}{2} + i 2 \pi M \xi}, \quad \gt= \frac{\hbar}{m} \frac{2\pi}{L^2} t, \quad \xi = \frac{(x'-x)}{L},
\end{equation}
evaluated at the boundary of the convergence domain $\im \gt<0$. Notice that a naive limit $L\to\infty$ of Eq.~\eqref{e_propa_circle_formal} leads to the propagator on the line, though. Also, on a circle the free propagator is periodic with period $2 \frac{L^2}{2\pi} \frac{m}{\hbar}$ in $t$ (equivalently of period $2$ in $\gt$), which is absent in the limit $L\to\infty$.

Still, physics leads us to believe that the free propagator is a definite quantity and therefore Eq.~\eqref{e_propa_circle_formal} is not a good starting point for its evaluation. If this were not so, then there will be no hope that any simulation of one-particle dynamics, classical or quantum, can be used to conclude something about the physics at finite $L$ and henceforth about the line limit. As such, the free propagator on the circle should be obtained through a different approach, which is the starting point for Sec.~\ref{s_propa_circle_finite_N} and Sec.~\ref{s_propa_circle_continuum}. Once this propagator is obtained in a closed form, then it is possible to combine Eq.~\eqref{e_propa_infinitesimal_circle} and Eq.~\eqref{e_path_integral_circle_propa} to formally solve the dynamics of generic wave functions Eq.~\eqref{e_time_evo_wave_f_circle}. In a simulation, only a finite amount of points can be sampled, and therefore we can only provide a numerical algorithm which solves for a finite number of position values. Concretely, the propagator will involve only a finite amounts of positions, as it is the case in a experiment with detectors. However, the logic tying the time evolution of the wave function to the propagator is unchanged. If the number of sampled points is sufficiently large, then we can use the result of the simulation to draw conclusions about the physics on the continuum. The algorithm of quantum computation we will review in the next section will provide as readout the time-evolved wave function for a particle on a circle discretized into a finite amount of points.

\section{Review of the Digital Quantum Simulation of One-Particle Dynamics}\label{s_rev_algo}

In this section we briefly review an algorithm of quantum computation which simulates one-particle dynamics in one-dimensional space \cite{MST2024}. 

The algorithm involves $n$ qubits $\ket{j_i} \in \cbr{\ket{0}, \ket{1}}$, which can be combined into $N_n= 2^n$ states labeled by the integers 
$\cbr{0, 1, \ldots, 2^n -1}$, also written in terms of
\begin{equation}\label{eq_convention_bits}
j = \overline{j_1 j_2 \ldots j_n} = j_1 2^{n-1} + j_2 2^{n-2}+\ldots+ j_n.    
\end{equation}
The notation also carries on for the states $\ket{j}$. These states are naturally put into correspondence with $N_n$ position eigenstates on the circle where the one-dimensional particle lives. The eigenvalues are mapped as
\begin{equation}
    \frac{x_J}{L} = -\half + \frac{j}{2^n} = \frac{J}{N_n} \in \left[-\half, \half \right), \quad J \in \cbr{-N_n/2 , -N_n/2 + 1, \ldots, N_n/2-1}.
\end{equation}
Here we have introduced the $J$'s, which are nothing more than the $j$ shifted by $N_n/2$. Since the $J$'s are nothing more than the positions in adimensional units, the momenta which are captured by the discretization are not the whole set of integers, but only a part:
\begin{equation}
    K_M = \frac{L}{N_n} \kappa_M = \frac{2\pi}{N_n} M, \quad M\in \cbr{-N_n/2 , -N_n/2 + 1, \ldots, N_n/2-1}. 
\end{equation}
In this regard, discretization of space provides a cut-off of larger momenta. By construction the set of allowed $J$'s and $K$'s is symmetric around $0$.

The set $\cbr{\ket{j}}$ is a basis for the Hilbert space of the qubits. We can find its conjugate $\cbr{\ket{k}}$ through the (discrete) quantum Fourier Transform
\begin{equation}
\label{e_qFT_qubit}
\ket{j} = \frac{1}{2^{n/2}}\sum_{k=0}^{2^n-1} e^{-2\pi i j k/2^{n}} \ket{k}, \quad k\in \cbr{0, 1,  \ldots, N_n -1},
\end{equation}
in analogy with Eq.~\eqref{e_qFT_circle} of Sec.~\ref{s_rev_oneparticle}. The operator $\widehat{\mathcal{F}}$ mapping $k$'s into $j$'s and vice-versa can be realized in terms of $O(n^2)$ single-qubit and controlled operations \cite{r_2010nielsen_b_qc_qi} and it is at the heart of the simulation algorithm. As much as the $j$'s are shifted from the $J$'s, also the $k$'s are shifted from the $K$'s \cite{MST2024} 
\begin{equation} \label{eq_momentum_k}
K_M = \begin{cases}
k , &\text{ if }k  = \lbrace 0, \ldots, 2^{n-1}-1\rbrace,\\
k-2^{n}, &\text{ if }k = \lbrace 2^{n-1}, \ldots, 2^{n}-1\rbrace.
\end{cases}
\end{equation}

The one-particle wave function $\psi(x)$ on the circle of lenght $L$ is encoded into the register of $J$'s as
\begin{equation}\label{e_initial_state}
\ket{f_0}= \sum_{J=-N_n /2}^{N_n/2 -1} f^J \ket{J}, \quad f^J = \sqrt{\frac{L}{N_n}}\psi(x_J).
\end{equation}
Encoding such a wave function into a register of qubits is generically hard, but for specific functions one can devise efficient implementations. For example, one for Gaussian wave functions has been proposed by Kitaev and Webb \cite{kitaevwebb2009}.

As far as the protocol of time evolution is concerned, it is best to work directly in terms of the unitary matrices performing the infinitesimal evolution with respect to the kinetic and potential terms. This avoids introducing a time scale in order to make the Hamiltonian adimensional. The adimensional time is given as in Eq.~\eqref{e_adim_time}, $\gt = \frac{\hbar}{m} \frac{2\pi}{L^2} t$, and the evolution with respect to the kinetic term reads
\begin{equation}\label{e_free_hamiltonian}
    \mathcal{K}_{\gt} =  \sum_{M=-N_n/2}^{N_n /2-1} \exp\br{-i 2\pi M^2\frac{\gt}{2}}\pro{M}.
\end{equation}
Notice that, in contrast to Sec.~\ref{s_rev_oneparticle}, only $N_n$ momenta contribute to the matrix. As far as the potential is concerned, the unitary operator is
\begin{equation}\label{e_harmonic_potential}
    \mathcal{V}_{\gt} = \sum_{J = -N_n /2}^{N_n /2 -1} \exp\br{- i u_J \gt} \pro{J}, \quad u_J = \frac{m}{\hbar^2} \frac{L^2}{2\pi} V(x_J).
\end{equation}
The equivalent of Eq.~\eqref{e_propa_infinitesimal_circle} is given by the Trotter expansion of the time evolution. Let $\mathcal{N}$ be the number of Trotter steps, and $\epsilon = \gt/\mathcal{N}$ the "infinitesimal" simulation time. The final state
\begin{equation}\label{eq_time_evo_circuit}
\includegraphics[scale=1]{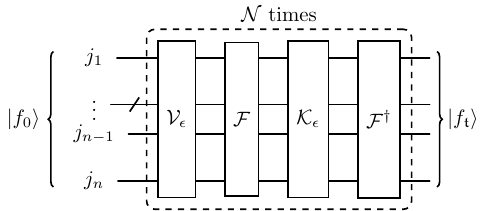}
\end{equation}
will approximate the wave function on a circle with error $O(\epsilon)$, assuming that perfect gates are used to implement the time-evolution. As far as the kinetic evolution is concerned, only $n^2$ perfect gates are required and an explicit gate construction is given in \cite{MST2024}. As far as the potential is concern, it follows the same considerations as the initial state preparation. 

Eq.~\eqref{eq_time_evo_circuit} is nothing more than the discrete version of Eq.~\eqref{e_path_integral_circle_propa}, written in terms of gates, instead of matrix products. The final state
\begin{equation}\label{e_path_integral_finite_propa}
    f^{J'}(\gt) = \sum_{J= -N_n/2}^{N_n /2 -1} D_{J}^{J'} (\gt) f_J = \sum_{J_\mathcal{N}, \ldots, J_2, J_1}  \mathcal{D}_{J_{\mathcal{N}}}^{J'}(\epsilon) e^{-i u_{J_{\mathcal{N}}} \epsilon}\mathcal{D}_{J_{\mathcal{N}-1}}^{J_{\mathcal{N}}}(\epsilon) e^{-i u_{J_{\mathcal{N}-1}} \epsilon}\ldots e^{-i u_{J_2} \epsilon} \mathcal{D}_{J_1}^{J_2}(\epsilon) e^{-i u_{J_1} \epsilon} \mathcal{D}_{J}^{J_1}(\epsilon) f^{J}
\end{equation}
is the outcome of the simulation. In other words, any measurement at the end of the computation will yield a state $\ket{J'}$ with probability $\norm{f^{J'}(\gt)}$. Tomography of the qubit register in the computational basis will reproduce the one-particle wave function truncated at finite $N_n$ and $L$.

Studying the latter equation is helpful in understanding the outcome of the quantum simulation. Since the potential and initial wave functions are given as input, the only quantity left to study is the free propagator
\begin{equation}\label{e_propa_finite_simulation}
    \mathcal{D}_{J}^{J'}(\gt) = \braket{J'| \mathcal{K}_{\gt} |J} = \frac{1}{N_n} \sum_{M=-N_n /2}^{N_n/2 -1} \exp\br{-i 2\pi M^2\frac{\gt}{2} + i \frac{2\pi}{L} M \frac{(J'-J)}{N_n}}.
\end{equation}
This free propagator is not the only possible definition of kinetic term. It is natural in the context of one-particle dynamics. In Appendix~\ref{a_adimensional_free_hamiltonian} we study another kinetic Hamiltonian, which may be more fitting in the case of abstract qubit dynamics. 

In contrast to Eq.~\eqref{e_propa_circle_formal}, the number of terms in the latter equation is finite. This allows for an analytic computation of the free propagator at finite $N_n$, to which is dedicated Sec.~\ref{s_propa_circle_finite_N}. Sec~\ref{s_propa_circle_continuum} is instead dedicated to the limit $N_n \to \infty$, leading to the propagator on the continuous circle.

\section{Derivation of the Free Propagator on a Circle Sampled at $N$ Equally Distanced Points}\label{s_propa_circle_finite_N}

In this section, we address the computation of the free propagator of Eq.~\eqref{e_propa_finite_simulation} for generic but finite $N$, not equal to $2^n$ necessarily. However, in order to obtain a swift generalization to the case $L\to\infty$ it is not optimal to start from Eq.~\eqref{e_propa_finite_simulation}, instead we proceed through different lines.

Let us imagine an ideal experiment in which we simulate the dynamics of one particle on a continuous circle of length $L$. As in Sec~\ref{s_rev_oneparticle}, $\ket{x}$ is an initial state of a particle localized at $x$. Let us assume that us, experimenters, can only detect the particle through a series of $N$ equally spaced detectors. At time $t$, when all detectors are switched on simultaneously, as part of the measurement process only one detector will activate; let it be $J$. By convention, if $N$ is even, the $J$'s are labeled $\cbr{-N/2, \ldots, N/2-1}$, while if $N$ is odd we have instead $\cbr{-(N-1)/2, \ldots, (N-1)/2}$. If detector $J$ activates, then we can conclude that the particle at time $t$ is found within $x_J \pm \ell/2$, where $\ell = L/N$ is the experimental precision -- or in other words the lattice spacing. Since the collection of $J$'s completely exhausts the outcomes of a measurement on a circle -- within experimental error, -- we can say that $\ket{J}$ is rightfully a quantum state, and is defined as in Sec.~\ref{s_rev_algo}. In assuming perfect detectors, measuring $J$ implies that the true position can not be found outside an interval of length $\ell$ centered around $x_J$, i.e. the states $\ket{J}$ are not overlapping. This allows writing
\begin{equation}\label{eq_continuous_discrete_states_circle}
    \ket{J} = \int_{x_J \pm \ell/2} \dd x\, \sqrt{\frac{N}{L}}\ket{x},
\end{equation}
stating a correspondence between discrete states $\ket{J}$ and continuous states $\ket{x}$. For any finite $N$, the correspondence is not one-to-one, though it becomes formally so in the limit $N\to\infty$. 

Following the ideal experiment, we can ask the question: if a particle initially localized at $\ket{x}$ is evolved for time $t$ under the kinetic Hamiltonian $K_L$ (containing infinitely many discrete contributions), what is the probability amplitude of detecting $\ket{J'}$? It is given by the ``mixed'' propagator
\begin{equation}\label{eq_propa_mixed_finite}
\begin{aligned}
    \braket{J'| e^{- i K_L t/\hbar} |x} &=\br{ \sum_{M'= -N/2}^{N/2-1} \frac{e^{i K_{M'}J}}{\sqrt{N}} \bra{K_{M'}}} e^{-i K_L t/\hbar} \br{\sum_{M\in \integers} \frac{e^{i \kappa_M x}}{\sqrt{L}}\ket{\kappa_M}}\\
    &=\sum_{M= -N/2}^{N/2-1} \frac{1}{\sqrt{L N}} \exp\br{- i 2\pi  M^2 \frac{\gt}{2} + i 2\pi M \br{\frac{J'}{N}-\frac{x}{L}}}.
    \end{aligned}
\end{equation}
This result is given for $N$ even, for brevity. The case of $N$ odd only entails a shift in the sum. Fundamentally, the latter identity holds because the states $\ket{M}$ for $M\in\cbr{-N/2, \ldots, N/2 -1}$ label eigenvalues of momenta in the basis $J$ and $x$, which only differ by a fixed constant $L/N$. Thus, if we consider simply $N$ points, all the momenta except those labeled $M\in\cbr{-N/2, \ldots, N/2 -1}$ are cut off by the finite lattice spacing $\ell$. Therefore for any finite $N$, the free propagator in the mixed representation Eq.~\eqref{eq_propa_mixed_finite} is a finite quantity. For example, if $\gt=0$ and $x= x_J$ for a certain $J$, then the propagator evaluates at
\begin{equation}
    \braket{J'| e^{- i K_L t/\hbar} |x_J}\rvert_{\gt = 0} = \sqrt{\frac{N}{L}} \krdel{J'}{J}.
\end{equation}
This $\sqrt{N}$ increase is characteristic of the propagator, and it must be so since proabability is conserved
\begin{equation}\label{e_mixed_propa_conservation_probability}
    \begin{aligned}
        1&= \sum_{J'=-N/2}^{N/2} \frac{L}{N} \norm{\sum_{M= -N/2}^{N/2-1} \inv{\sqrt{LN}} \exp \br{- i 2\pi  M^2 \frac{\gt}{2} + i 2\pi M \br{\frac{J'}{N}-\frac{x}{L}}}}\\
        &= \int_L \dd x\, \norm{\sum_{M= -N/2}^{N/2-1} \inv{\sqrt{LN}} \exp \br{- i 2\pi  M^2 \frac{\gt}{2} + i 2\pi M \br{\frac{J'}{N}-\frac{x}{L}}}}.
    \end{aligned}
\end{equation}

The propagator Eq.~\eqref{eq_propa_mixed_finite} is a highly nontrivial function of the simulation time $\gt$. First of all, it is a function periodic on the circle with period $L$, but also periodic in the simulation time $\gt$ with period $2$. As such, we will only consider $\gt \in \sbr{-1,1}$. If $x = x_J$ for some $J$, at time $\gt=0$, the particle will be localized at $x_J$, as previously seen. One would be led to think that this propagator is uniformly distributed for nonzero times, as the propagator on the line Eq.~\eqref{e_propa_line_limit}. It is not so, though: one can readily check that for $\gt =1$ that the propagator is nonzero only for $J' = J - N/2$. 

The outcome of this preliminary investigation is that the free propagator on the circle is not necessarily uniformly distributed for all $\gt$, at any given simulation time, but there are only a set of "allowed outcomes" of a position measurement, which are a subset of the whole set of sampled points. These positions can be obtained though an heuristic method, which will be presented now. These allowed positions are therefore linked to the absolute value of the free propagator, i.e. the transition probability under kinetic motion. In order not to clutter the notations, we will use the adimensional units already introduced in Eq.~\eqref{e_adim_time}:
\begin{equation}\label{e_adim_units}
    \gt= \frac{\hbar}{m} \frac{2\pi}{L^2} t, \quad \xi' = \frac{J'}{N}, \quad \xi = \frac{x}{L}, \quad \Delta\xi = \xi'-\xi.
\end{equation}
We will also show that given a set of allowed positions at some rational $\gt$, only some of them are truly measured by a position measurement. They will be henceforth called "physical outcomes".

We may imagine the propagator to be composed of interfering waves, each carrying a phase at a given displacement and time:
\begin{equation}\label{e_wave_phase}
    \varphi_M = - \kappa_M \br{(x'-x) - \frac{\hbar}{m}\kappa_M \frac{t}{2}} = - 2\pi M \br{(\xi'-\xi) - M \frac{\gt}{2}} = - 2\pi M \br{\Delta\xi - M \frac{\gt}{2}}.
\end{equation}
If we were to consider these waves as individual particles moving on a circle, the most probable positions are those for which the phases $\varphi_M$ vanish. We then find a set of "allowed" positions on the circle, given by
\begin{equation}\label{e_particle_interpretation}
    \Delta \xi_M = M \frac{\gt}{2}.
\end{equation} 

Because of the periodicity in the simulation time $\gt \in [-1, 1)$, we distinguish two cases:
\begin{enumerate}
    \item If $\gt$ is rational, i.e. $\gt = p/q$, where $p$ and $q$ are coprimes, then the set of most probable positions is given by
    \begin{equation}\label{e_allowed_q}
        \cbr{\Delta \xi} = \half \frac{\cbr{-q, -q+1, \ldots, 0, \ldots, q-1}}{q}.
    \end{equation}
    There are a total of $2q$ total displacements accessible at these simulation times. However, not all these are physical outcomes of a measurement, given $p$. In the following, we shall differentiate by the allowed displacements, which are only determined by $q$, and the physical ones, which depend on both $p$ and $q$.
    
    In this picture, the integer $p$ scales the velocities of the particles. We can define an effective velocity, $v_{e} = pM$ which dictates the motion for a reference time $1/q$. Therefore, the difference between the case $\gt = 1/q$ and $\gt = p/q$ is that in the latter case the particle is moving $p$ times faster than in the former.
    
    \item If $\gt$ is irrational, we can use a sequence of rationals to approximate it. For example, we can write $\inv{\pi} = \cbr{\frac{1}{3}, \frac{10}{31}, \frac{100}{314}, \ldots }$. Each time in the sequence is rational and therefore can be treated as in the former case. At each iteration, the number of allowed positions increases, implying that in the continuum limit the set of allowed points is the whole circle. Also the numerator increases with each term of the sequence, therefore the effective velocities will become greater and greater, implying that two infinitesimally close positions on the circle are not reached with continuity by the particle, but it may take an arbitrary large number of windings to move between the two. This phenomenon justifies that in the continuum limit, for $\gt$ irrational, the phase of the propagator can not be written as a continuous function of the displacement $\Delta \xi$.

    Crucial is the time $\gt = \epsilon$, where $\epsilon$ is an irrational number smaller than any $1/N$. For this specific time, all positions on the circle are allowed and they can indeed be reached with continuity. This will be the only case surviving the limit $L\to\infty$, yielding Eq.~\eqref{e_propa_line_limit}.  
\end{enumerate}

Therefore, we have reduced the propagator to the study at $\gt = p/q$. It is also not surprising that the propagator evaluated at $\gt' = -\gt$ will have the same allowed points as $\gt$, but opposite phase. This is due to the reversal of particle velocities, or in other words the orientation of the circle.

If $\gt = p/q$, the free propagator at the allowed positions $\cbr{\Delta\xi}$ Eq.~\eqref{e_allowed_q} will be given by $2q$ interfering waves. The correct interference pattern is retrieved in two scenarios: in the limit $N\to\infty$ and when $N$ is a multiple of $q$. Therefore out of the set of allowed points, only a subset of them will contribute constructively to the propagator, and these will be the physical outcomes of a position measurement in the ideal experiment of a free evolution for time $\gt$. We recall that in summing
\begin{equation*}
    e^{i \varphi_{M}} + e^{i \varphi_{M'}} = 2 \cos\br{\frac{\varphi_{M}-\varphi_{M'}}{2}} \exp\br{i \frac{\varphi_{M}+\varphi_{M'}}{2}},
\end{equation*}
the interference is constructive if $\varphi_{M}-\varphi_{M'}= 2\pi \integers$ and destructive if $\varphi_{M}-\varphi_{M'} = \pi + 2\pi \integers$. In our case
\begin{equation}
    \varphi_{M}-\varphi_{M'} = 2\pi (M'-M) \Delta \xi - 2\pi (M'^2 - M^2) \frac{\gt}{2}.
\end{equation}

At this moment, we have to differentiate between two cases: $q$ even and $q$ odd. Since the analysis is the same in both cases, we will outline the derivation for even $q$. This latter case  is the most relevant for quantum computation, because the number of states given by $n$ qubits is always even.

If $q$ is even, $q= 2Q$, then the set of coprime $p$'s is made of odd numbers which do not divide $Q$. The set of allowed $M$'s in Eq.~\eqref{e_allowed_q} is $\cbr{-2Q, -2Q+1, \ldots, 2Q-1}$, up to periodic shifts. Since $\Delta \xi=0$ is always an allowed position, we study first the phase at this special point:
    \begin{equation*}
        \Delta \xi = 0: \quad 2\pi (M'^2 - M^2) \frac{p}{4Q} = 2\pi \integers.
    \end{equation*}
    Since the number of $M$'s is $4Q$, we can find $2Q$ couples, which we choose simply as
    \begin{equation*}
        M' = -2Q + s, \quad M = s, \quad s = \cbr{0, 1, \ldots, 2Q-1}.
    \end{equation*}
    For any $s$, the phase difference reads
    \begin{equation*}
        \Delta \xi = 0: \quad 2\pi (Q-s)p = 2\pi \integers,
    \end{equation*}
    which is always true because of the factor $2$. Therefore $\Delta\xi=0$ is always allowed because of constructive interference. In considering the other $\Delta \xi$'s different from zero, the physical positions are given by 
    \begin{equation*}
        \Delta \xi = \half\frac{M''}{2Q}: \quad 2\pi (M'-M)\frac{M''}{4Q} =- \pi M'' = 2\pi \integers, \quad M'= -2Q+s,\, M=s.
    \end{equation*}
    The only allowed values are $M''$ even, i.e. out of all the $4Q=2q$ points, only half of them are really reached by propagation. In conclusion the physical outcomes of a free propagation in this case are
    \begin{equation}\label{eq_physical_even}
        \gt = \frac{p}{q},\; q\; \text{even,}\; p\; \text{odd coprime:} \quad \cbr{\Delta\xi }_{\text{ph}} = \frac{1}{q}\cbr{-\frac{q}{2}, -\frac{q}{2}+1, \ldots, \frac{q}{2}-1}.
    \end{equation}

If $q$ is odd, a similar reasoning leads to
\begin{equation}\label{eq_physical_odd}
\begin{gathered}
        \gt = \frac{p}{q},\; q\; \text{odd,}\; p\; \text{odd coprime:} \quad \cbr{\Delta\xi}_{\text{ph}} = \frac{1}{2q}\cbr{-q, -q + 2, \ldots -1, 1, \ldots, q-1},\\
        \gt = \frac{p}{q},\; q\; \text{odd,}\; p\; \text{even coprime:}  \quad \cbr{\Delta\xi}_{\text{ph}} = \frac{1}{2q}\cbr{-q+1, -q+3, \ldots, 0, \ldots, q}.
\end{gathered}
\end{equation}

\begin{figure*}
\centering
\begin{subfigure}{0.3\textwidth}  
\includegraphics[scale=1]{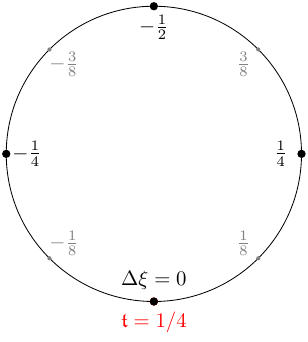}
\caption{}
\end{subfigure}\hfill
\begin{subfigure}{0.3\textwidth}  
\includegraphics[scale=1]{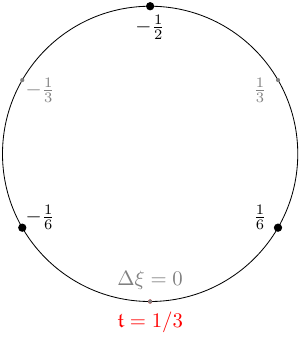}
\caption{}
\end{subfigure}\hfill
\begin{subfigure}{0.3\textwidth}  
\includegraphics[scale=1]{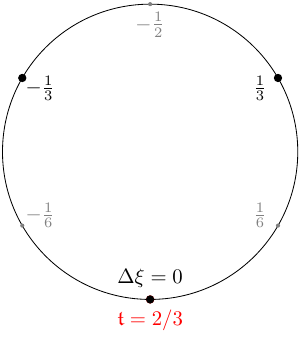}
\caption{}
\end{subfigure}
\caption{Some examples of the physical displacements (black dots) measurable out of the allowed ones (black and grey dots) for the rational times $\gt= 1/4$ (a), $\gt =1/3$ (b) and $\gt = 2/3$ (c). Out of the allowed positions, the physical ones are those for which the waves interfere constructively.}
\label{f_circle_physical_positions}
\end{figure*}

The results of Eq.~\eqref{eq_physical_even} and Eq.~\eqref{eq_physical_odd} and the difference between allowed and physical positions is better understood from examples. In Fig.~\ref{f_circle_physical_positions} we present three cases: $\gt = 1/4$, $\gt = 1/3$ and $\gt = 2/3$. 

Through this argument it is possible to find the physical outcomes of free propagation on the circle. All physical outcomes are equally probable, with probability $1/(L q)$, hence the modulus of the propagator is found. For the phase, it is only rightfully defined at the physical positions. It can be easily found numerically, but much insight can be obtained if we manage to find and analytic continuation to some  function of $\Delta\xi$.

Generally, determining the phase means to sum, given $\gt=p/q$ and the physical displacements, the quantity 
\begin{equation}
    \half \inv{\sqrt{q}}\sum_{M=-q}^{q-1} \exp \br{-i 2\pi \frac{p M^2}{2q} + i 2 \pi M \Delta \xi}.
\end{equation}
Even the simplest case of $\Delta\xi=0$, for $\gt = 1/q$ and $q$ even, is highly nontrivial because of the finite sum. Deferring the proof to Appendix~\ref{a_technical}, it can be shown to be, for any positive integer $q$:
\begin{equation}\label{eq_phase_0_q_even}
    \half \inv{\sqrt{q}}\sum_{M=-q}^{q-1} e^{-i 2\pi \frac{M^2}{2q}} = \inv{\sqrt{q}}\sum_{M=-q/2}^{q/2-1} e^{-i 2\pi \frac{M^2}{2q}} =\inv{\sqrt{q}}\sum_{M=0}^{q-1} e^{-i 2\pi \frac{M^2}{2q}} = e^{-i \pi/4},
\end{equation}
coinciding with the relative phase appearing in the free propagator on the line, which in the latter case is a byproduct of Gaussian integration. 

For all the values $\gt = 1/q$, $q$ even, we conjecture that the phase associated to two physical positions on the circle $\xi'$ and $\xi$ is
\begin{equation}
    S^{\xi'}_{\xi}(\gt) = \operatorname{sign}(\gt)\br{\pi \frac{(\xi'-\xi)^2}{\abs{\gt}} - \frac{\pi}{4}}, \quad (\xi'-\xi)\in [-1/2, 1/2), \quad \gt = 1/q, \quad q\in \integers.
\end{equation}
In mapping negative to positive $\gt$, we must take into account that the velocities of the waves are mapped to the opposites, and therefore the propagator at negative $\gt$ is the complex conjugate of the one at positive $\gt$. If one is willing to find the phase for generic $p$, it can be found by repeated applications of Eq.~\eqref{e_convo_propagator}, upon converting integrals into matrix multiplications.

We now consider the sequence $\cbr{1, 1/2, 1/3, 1/4, \ldots}$ and define $\epsilon$ as the "infinitesimal" simulation time such that it is smaller than any entry of the sequence. Then $\epsilon$ will be irrational and therefore the number of points is the whole continuum on the circle. This time can only be simulated if $N\to\infty$. If $N$ is finite, the smallest time one can utilize for the simulation, utilizing a uniform distribution of points for free propagation, is $\gt = 1/N$. As such, on a quantum computer with $n$ qubits, the time $\gt_n = 2^{-n}$ is a natural simulation time for the Trotter step in 
Eq.~\eqref{eq_time_evo_circuit}.

We conclude this section providing the formula for the free propagator with finite $N$ and $\gt = 1/q$, where $q$ is a divisor of $N$. For such values the propagator is exact and the probabilities will depend on the ratio $N/q$. For the points $x_J$ we have
\begin{equation}
\begin{aligned}\label{eq_propa_circle_exact_finite}
    \braket{J'| e^{- i K_L t/\hbar} |x_J}\rvert_{\gt = \inv{q}} 
    &=\sum_{M= -N/2}^{N/2-1} \frac{1}{\sqrt{L N}} \exp\cbr{- i \frac{2\pi}{\gt} M^2  + i (2\pi M) \br{\frac{J'}{N}-\frac{x_J}{L}}}\\
    &= \frac{1}{\sqrt{L}} \sum_{M= -q/2}^{q/2-1} \sqrt{\frac{N}{q}} \krdel{J'- N(\xi'-\xi)_M}{J} \exp\br{i S_{J/N}^{J'/N}(\gt)},\\
    \norm{\braket{J'| e^{- i K_L t/\hbar} |x_J}\rvert_{\gt = \inv{q}} }&=\frac{N}{L} \sum_{M=-q/2}^{q/2-1} \inv{q} \krdel{J'- N(\xi'-\xi)_M}{J}.
    \end{aligned}
\end{equation}
Here the $(\xi'-\xi)_M$ label the $q$ physical positions of Eq.~\eqref{eq_physical_even}. For $q$ odd, the only thing that changes are the extrema of the summation and the physical displacements given by Eq.\eqref{eq_physical_odd}.

This latter form of the propagator only includes $q$ terms in the sum, and all physical displacements appear with uniform probability. It is from this object that we can take the continuum limit. It can also be checked that from Eq.~\eqref{eq_propa_circle_exact_finite}, the conservation of total probability Eq.~\eqref{e_mixed_propa_conservation_probability} is always guaranteed and is independent of $N$ -- provided that it is a divisor of $q$. If $N$ is not a multiple of $q$, then some physical displacements are not corresponding to any integer sequence $(J'-J)/N$. When $N$ is multiple of $q$, allowed position will correspond to $(J'-J)/N$.

\section{The Free Propagators on a Continuum Circle and the Line Limit}\label{s_propa_circle_continuum}

In this Section, we address the continuum limit of Eq.~\eqref{eq_propa_circle_exact_finite} and the line limit leading to Eq.~\eqref{e_propa_line_limit}. 

The technical problem in taking the limit $N\to\infty$ is that the infinite but discrete basis of $\ket{J}$ must be connected with the compact and continuous basis of $\ket{x}$. Starting from the mixed representation, at finite $N$, Eq.~\eqref{e_propa_finite_simulation} becomes 
\begin{equation}\label{eq_propa_finite_Jbasis}
    \braket{J'| e^{- i K_L t/\hbar} |J}\rvert_{\gt = \inv{q}} 
    = \sum_{M= -q/2}^{q/2-1} \inv{\sqrt{q}} \krdel{J'- N(\xi'-\xi)_M}{J} \exp\br{i S_{J/N}^{J'/N}(\gt)},
\end{equation}
which is exact as long as $N$ is multiple of $q$. On the discrete basis, probability is conserved. On the other hand, in passing from the mixed propagator to Eq.~\eqref{e_propa_circle_formal} we expect to find divergencies. Let us denote by $\braket{x'|e^{-i K_L t_\epsilon /\hbar}|x}_R $ the ``regularized'' free propagator, which is of dimension of inverse length $\ul^{-1}$, in light of Eq.~\eqref{eq_continuous_discrete_states_circle}. This should not be seen as a problem, but rather as a reassurance, since also the free propagator on the line Eq.~\eqref{e_propa_line_limit} has the same dimension. 

Concerning the large $N$ behavior, the free propagator in the mixed representation Eq.~\eqref{eq_propa_circle_exact_finite} will diverge as $\sqrt{N}$ only in correspondence of the $(J'-J)/N$ equal to some of the physical positions of Eq.~\eqref{eq_physical_even}. For other differences $(J'-J)/N$, the propagator will be vanishing. This suggests that the free propagator in the limit will be constituted by isolated peaks, for $\gt$ rational. On the other hand, for $\gt$ irrational, all points on the circle will be physical positions and will appear with uniform probability.

Stemming from these considerations, the continuum limit that allows the transition from the propagator in the mixed representation to the regulated one calls for a mathematical object which is divergent whenever $\gt$ is rational and smoothens to a uniform probability density whenever $\gt$ becomes irrational. We will name such distribution $\sigma(x',x)$, formally defined as
\begin{equation}\label{e_sigma_defi}
    \sigma\br{x_{J'}, x_{J}} \defi \Lim{N}{\infty} \sqrt{\frac{N}{L}} \krdel{J'}{J}, \quad x_{J} = L J/N.
\end{equation}
Its fundamental properties will be derived following the physical interpretation of the free propagator, that is: as a probability density and as the kernel of time evolution on a circle. Moreover, it admits a well-known line limit Eq.~\eqref{e_propa_line_limit}. 
Also, the $\sigma$ Eq.~\eqref{e_sigma_defi} has a connection to Eq.~\eqref{eq_continuous_discrete_states_circle}: it allows for connecting, when $N\to\infty$, the state $\ket{J}$, infinite but discrete, to $\ket{x_J}$, continuous but compact. 

With this distribution, we can write the regularized free propagator on the circle at rational times
\begin{equation}\label{e_propa_continuum_circle_rationals}
    \mathscr{D}_{x}^{x'}(\gt)\rvert_{\gt = 1/q} \coinc \braket{x'|e^{-i K_L t/\hbar}|x}\rvert_{\gt = 1/q} = \inv{\sqrt{L}}\sum_{M= -q/2}^{q/2-1} \sqrt{\frac{1}{q}} \sigma\br{x' - L(\xi'-\xi)_M, x} \exp\br{i S_{\xi}^{\xi'}(\gt)}.
\end{equation}
The first property of the $\sigma$ is derived in such a way to guarantee that for infinitesimal simulation times $\gt = \epsilon$, i.e. $q\to\infty$, we retrieve the propagator on the line. The reason for this is that $\epsilon = \frac{\hbar}{m} \frac{2\pi}{L^2} t_\epsilon$ 
is always infinitesimal for every $t_\epsilon$ in the limit $L\to\infty$. Therefore as long as we consider $\gt= \epsilon$, the simulation is giving information about propagation on the continuum line. And since we know what is the behavior in the continuum line, Eq.~\eqref{e_propa_line_limit}, we can get advantage from this. For any sequence $\cbr{f_M}$ sampled from a function $f(x)$ on the circle, we require that 
\begin{equation}
    \Lim{q}{\infty} \sum_{M=-q/2}^{q/2} \inv{\sqrt{q}} \sigma\br{x'- \frac{L}{q} M, x} f_M = \Lim{q}{\infty}\sum_{M=-q/2}^{q/2} \inv{\sqrt{q}} \sigma\br{x'-x, L \frac{M}{q}} f_M = \inv{\sqrt{\abs{L/q}}} f\br{x'-x}.
\end{equation}
This relation is useful in connecting sequences to continuous functions, exactly what is needed to transform the isolated peaks of Eq.~\eqref{e_propa_continuum_circle_rationals} into
\begin{equation}\label{e_propa_continuum_circle_infinitesimal}
    \mathscr{D}_{x}^{x'}(\epsilon) = \braket{x'|e^{-i K_L t_\epsilon /\hbar}|x}_R = \sqrt{\frac{m}{2\pi\hbar t_\epsilon}}\exp\br{i S_{x'/L}^{x/L}\br{\frac{\hbar}{m}\frac{2\pi}{L^2}t_\epsilon} }=\sqrt{\frac{m}{i 2\pi\hbar t_\epsilon}}e^{\frac{i}{\hbar} m \frac{(x'-x)^2}{2 t_\epsilon}}.
\end{equation}
Notice that the dependence on $L$ is completely absent, as it should on the continuum, where no explicit length scale is provided, unless details on the wave function are given. 

At irrational times, as in the latter case, the outcome of a position measurement is a point on the continuous circle, and probability is not strictly conserved. The physically meaningful content is that the probability density is finite. Conversely, when $\gt$ is rational, the probability density is divergent, but since the outcomes of a measurement position are isolated points on the circle, probability should anyhow be conserved. As such, to ensure conservation of probability at $\gt = 1/q$ in Eq.~\eqref{e_propa_continuum_circle_rationals} we require
\begin{equation}\label{e_sigma_delta}
    \sum_{M' = -q/2}^{q/2-1} \sigma\br{x' - L(\xi'-\xi)_{M'}, x}\sigma \br{x' - L(\xi'-\xi)_{M}, x} = \delta\br{(x'-x) - L(\xi'-\xi)_{M}}, \quad \forall M\in\cbr{-q/2, \ldots, q/2-1},
\end{equation}
justifying the interpretation of $\sigma$ as the "square root" of the Dirac $\delta$ and as such has engineering dimensions of $\ul^{-1/2}$. Conservation of total probability is ensured when integrating over one period of the circle.

Irrespective of $\gt$, the propagator is the kernel of time evolution and Eq.~\eqref{e_time_evo_wave_f_circle} must be verified. It is certainly true for any periodic wave function -- physical states on $\mathcal{H}_{L}^{\text{P}}$ -- if for infinitesimal simulation times
\begin{equation}
   \inv{\sqrt{L}} e^{-i \frac{\hbar }{2m} \kappa^2(t+t_\epsilon) + i \kappa x'} = \int_L \dd x\, \braket{x'|e^{-i K_L t_\epsilon /\hbar}|x}_R \inv{\sqrt{L}} e^{-i \frac{\hbar }{2m} \kappa^2 t + i \kappa x}, \quad \forall \kappa \in \momentumlattice{L}.
\end{equation}

At least two proofs can be devised, starting respectively from Eq.~\eqref{e_propa_continuum_circle_infinitesimal}, as in \cite{feynman1948path}, or Eq.~\eqref{e_propa_continuum_circle_rationals}. We report here the last one, for which requiring $\kappa \in \momentumlattice{L}$ is essential. In order to perform the integration over the circle, we give another property regarding the $\sigma$: if $f(x)$ is a function periodic on the circle, then
\begin{equation}
    \frac{1}{\sqrt{L}}\int_L \dd x\, \sigma(x', x) f(x) = f(x').
\end{equation}
From the latter it follows straightforwardly: 
\begin{align*}
    &\inv{\sqrt{L}}\int_L \dd x\, \sum_{M= -q/2}^{q/2-1} \sqrt{\frac{1}{q}} \sigma\br{x' - L(\xi'-\xi)_M, x} \exp\br{i S_{\xi'}^{\xi}(\gt)} \inv{\sqrt{L}} e^{-i \frac{\hbar }{2m} \kappa^2 t + i \kappa x} \\
    &= \inv{\sqrt{L}}\sum_{M= -q/2}^{q/2-1} \sqrt{\frac{1}{q}} \exp\br{i S_{\xi'}^{\xi}(\gt)} e^{-i \frac{\hbar }{2m} \kappa^2 t + i \kappa \br{x' - L(\xi'-\xi)_M}}\\
    &= \inv{\sqrt{L}} e^{-i \frac{\hbar }{2m} \kappa^2 t + i \kappa x'} \sum_{M= -q/2}^{q/2-1} \sqrt{\frac{1}{q}} \exp\br{i S_{\xi'}^{\xi}(\gt)}e^{ -i \kappa L(\xi'-\xi)_M}\\
    &= \inv{\sqrt{L}} e^{-i \frac{\hbar }{2m} \kappa^2 t + i \kappa x'} \sum_{M= -q/2}^{q/2-1} \sqrt{\frac{1}{q}} \exp\br{i \operatorname{sign}(q) \br{\pi q (\xi'-\xi)^2_M - \frac{\pi}{4}}}e^{ -i \kappa L(\xi'-\xi)_M}\\
    &\underset{q>0}{=} \inv{\sqrt{L}} e^{-i \frac{\hbar }{2m} \kappa^2 t + i \kappa x'} \inv{\sqrt{i}}\sum_{M= -q/2}^{q/2-1} \sqrt{\frac{1}{q}} e^{i \pi \frac{M^2}{q} -i \kappa L \frac{M}{q}}\\
    &\underset{\forall \kappa\in\momentumlattice{L}}{=}\inv{\sqrt{L}} e^{-i \frac{\hbar }{2m} \kappa^2 t + i \kappa x'} \inv{\sqrt{i}} e^{-i \pi \frac{\kappa^2 L^2}{(2\pi)^2 q} + i \frac{\pi}{4}} = \inv{\sqrt{L}} e^{-i \frac{\hbar }{2m} \kappa^2 (t+t_q) + i \kappa x'},\quad t_q = \frac{L^2}{2\pi}\frac{m}{\hbar}\inv{q}.
\end{align*}
We believe that the proof of the last sum can be proven along the lines of Appendix~\ref{a_technical}.

In conclusion of this section, the free propagator on the circle we constructed for rational Eq.~\eqref{e_propa_continuum_circle_rationals} and infinitesimal Eq.~\eqref{e_propa_continuum_circle_infinitesimal} simulation times is such that
\begin{enumerate}
    \item It can be interpreted as a probability density amplitude, which is uniform for $\gt$ irrational, but not normalized when integrated on a circle. On the other hand, when $\gt$ is rational, overall probability is conserved but the probability density amplitude is a discontinuous function of isolated peaks.
    \item Yields the propagator on the continuous line either by considering infinitesimal times or by taking $L\to\infty$.
    \item Is the proper kernel of time evolution for all the elements in the basis of periodic wave functions on the circle.
\end{enumerate}
Equipped with these three properties, we can claim that such propagator is indeed the propagator on the circle, which can be simulated exactly on a digital quantum computer for times $\gt = 1/2^{n}$ and multiples thereof.

\section{Numerical Investigation of Time Evolution}\label{s_numerical}

In this section, we make use of the results of Sec.~\ref{s_propa_circle_finite_N} to simulate the outcome of the digital quantum simulation outlined in Sec.~\ref{s_rev_algo}.

The outcome of the quantum algorithm Eq.~\eqref{eq_time_evo_circuit} is a final state $\ket{f_\gt}$ which can be subjected to tomography, i.e one can measure onto the qubit computational basis -- the one of the Pauli $Z$ operator -- and reconstruct the probabilities $\norm{f^J (\gt)}$ if a sufficient number of samples is collected. On the other hand, each measurement yields one $J$ as outcome, and therefore one can reconstruct a trajectory $J(\gt)$ by looking at the maximum of probability $\norm{f^J (\gt)}$, for a given state $\ket{f}$, as function of the simulation time $\gt$. 

We will be interested in simulating wave packets. On the continuum line, it is well known that to a wave packet representing a particle localised initially at $x_0$ with variance $1/\sigma^2$ and peak momentum $k_0$, corresponds a position wave function
\begin{equation*}
    \psi^{\infty}_{x_0; k_0; \sigma}(x) = \frac{\sigma}{\sqrt{\pi}} \exp\br{-\frac{(x-x_0)^2}{2/\sigma^2} + i k_0 (x-x_0)},
\end{equation*}
and consequently a momentum wave function given by its Fourier Transform
\begin{equation*}
    \Psi^{\infty}_{x_0; k_0; \sigma}(k) = \inv{\sqrt{\pi} \sigma}\exp\br{-\frac{(k-k_0)^2}{2 \sigma^2} + i k (x-x_0)}.
\end{equation*}
On a circle, we discretize momenta and define the normalized wave packet corresponding to a particle initially localized at $x_0$ on the circle, with variance $1/\sigma^2$ and peak momentum $\kappa_0\in \momentumlattice{L}$, as
\begin{equation}\label{e_wave_packet_circle}
\begin{aligned}
     \psi^{L}_{x_0; \kappa_0; \sigma}(x) &= \inv{\sqrt{\theta\br{-i \frac{\kappa_0}{L\sigma^2};-i \frac{4\pi}{\sigma^2 L^2}}}} \sum_{\kappa \in \momentumlattice{L}} \frac{e^{i \kappa (x-x_0)}}{\sqrt{L}} e^{- \frac{(\kappa-\kappa_0)^2}{2\sigma^2}}\\
     \comment{
     &= \frac{\sigma \sqrt{L}}{\sqrt{2\pi \theta\br{-i \frac{\kappa_0}{L\sigma^2};-i \frac{4\pi}{\sigma^2 L^2}}}} \exp\br{-\frac{(x-x_0)^2}{2/\sigma^2} + i \kappa_0 (x-x_0)} \theta\br{-\frac{L\kappa_0}{2} - \frac{i}{2} \sigma^2 L (x-x_0); - \frac{i}{2\pi} \sigma^2 L^2}}
     &=\inv{\sqrt{L\theta\br{-i \frac{\kappa_0}{L\sigma^2};-i \frac{4\pi}{\sigma^2 L^2}}}}\theta \br{\frac{(x-x_0)}{L} - i \frac{\kappa_0}{L\sigma^2}; - i \frac{2\pi}{L^2 \sigma^2}}
     ,
\end{aligned}
\end{equation}
where $\theta$ is defined as in Eq.~\eqref{e_adim_time}. The normalization is readily found by considering the first equality. \comment{Quite amusingly, it is possible to extract the leading asymptotic behavior of $\theta\br{0;-i \frac{4\pi}{\sigma^2 L^2}}$ for large $L$, since it has to match the wave packet on the line:
\begin{equation*}
    \theta\br{0;-i \frac{4\pi}{\sigma^2 L^2}} \underset{L\gg 1}{\sim} \sqrt{\frac{L \sigma}{2}}\br{1 + O\br{\inv{L}}}.
\end{equation*}}

Our initial state is therefore given by Eq.~\eqref{e_initial_state} with the particular wave function Eq.~\eqref{e_wave_packet_circle}. Concerning the time evolution, first we study what happens with the free propagation Eq.~\eqref{e_free_hamiltonian}. Each plane wave $\ket{\kappa}$ is eigenvector of the free Hamiltonian, therefore it is immediate to write the time-evolved wave function of a free packet
\begin{equation}\label{e_time_evo_free_wave_packet_circle}
\begin{aligned}
     \psi^{L}_{x_0; \kappa_0; \sigma}(x; t) &= \inv{\sqrt{\theta\br{-i \frac{\kappa_0}{L\sigma^2};-i \frac{4\pi}{\sigma^2 L^2}}}} \sum_{\kappa \in \momentumlattice{L}} \frac{e^{i \kappa (x-x_0)}}{\sqrt{L}} e^{- \frac{(\kappa-\kappa_0)^2}{2\sigma^2} - i \frac{\hbar}{m}\kappa^2 \frac{t}{2}}\\\comment{
     &= \frac{\sigma \sqrt{L}}{\sqrt{2\pi \br{1+ i \sigma^2  t \hbar/m } \theta\br{0;-i \frac{4\pi}{\sigma^2 L^2}}}} \exp\br{\frac{-\frac{(x-x_0)^2}{2/\sigma^2} + i \kappa_0 \br{x-x_0 -\frac{\hbar}{2m} \kappa_0 t}}{1+ i \sigma^2  t \hbar/m}}\\
     &\quad \times\theta\br{\frac{-\frac{L\kappa_0}{2} - \frac{i}{2} \sigma^2 L (x-x_0)}{1+ i  \sigma^2  t \hbar/m}; -i \frac{{\sigma^2 L^2 /(2\pi)}}{{1+ i  \sigma^2  t \hbar/m}}}}
      &= \inv{ \sqrt{L\theta\br{-i \frac{\kappa_0}{L\sigma^2};-i \frac{4\pi}{\sigma^2 L^2}}}}\theta \br{\frac{(x-x_0)}{L} - i \frac{\kappa_0}{L\sigma^2}; - i \frac{2\pi}{L^2 \sigma^2 }+\gt}.
\end{aligned}
\end{equation}
In analogy to the case on the line \cite{CohenTannoudji}, one can define a semi-classical velocity $v_0 = \frac{\hbar}{2m} \kappa_0 t$ and a time-dependent packet variance
    $1/\sigma^2 (t) = \br{1+ \br{ \frac{\hbar}{m} t \sigma^2 }^2}/\sigma^2$.
The interpretation of the physics on a circle is somehow different than on the line.

As in the case of a line \cite{CohenTannoudji}, a free particle on a circle will initially delocalize and have a probablity density maximum traveling with velocity $v_0$ -- to the right if positive, to the left if negative. However, this dynamics competes with the time-evolved wave packet being periodic in $\gt$ with period $2$. In general, this periodicity is absent when the potential is nonzero, for the simple reason that plane waves of definite $\kappa$ are not eigenfunctions anymore. As such at late times, $v_0$ and $1/\sigma(t)$ are not coinciding with the peak velocity and the time dependent spreading, because in Eq.~\eqref{e_time_evo_free_wave_packet_circle} the Gaussian is dressed by the $\theta$. Because of periodicity, to the initial delocalization follows a ``re-localization'' as $\gt$ approaches 2. Of course, in the limit $L\to\infty$, any physical time $t$ will make the corresponding $\gt$ infinitesimal, therefore only the small $\gt$ dynamics is that which survives the continuum line limit. As a consequence, a particle on a line will spread forever with $1/\sigma(t)$ and have a probability density maximum localized at $x_t = x_0 + v_0 t$. Notice that both quantities are independent of $L$, as it should be in the line limit. 
\comment{
Eq.~\eqref{e_time_evo_free_wave_packet_circle} can be given an analytic form because of the simplicity of the free Hamiltonian. In general, when the potential is nonzero, one needs first to diagonalize the Hamiltonian and then express the eigenspectrum in position or momenta bases.}

Another strategy to compute the time evolved wave function, which works also in presence of potentials, is to employ a numerical method to find the time evolution at discrete times. Having in mind the setup of quantum computation, we consider $N_n = 2^{n}$ equally distanced points and partition the circle. We can use Eq.~\eqref{e_path_integral_circle_propa}, in absence of potential, provided that we specify the simulation time. We argued in Sec.~\ref{s_propa_circle_finite_N} that the propagator on the circle can not be faithfully represented for arbitrary $\gt$ with finite $N_n$, but only for simulation times which are of the form $p/q$, where $q$ is multiple of $N_n$. At these times, each physical position appearing in the free propagator can be detected with uniform probability. If the points are $2^n$, then the simulation times we can simulate without invalidating the results of Sec.~\ref{s_propa_circle_finite_N} are the multiples of $\gt_n = 2^{-n}$. Hence the smallest simulation time we probe in a simulation through Eq.~\eqref{e_path_integral_circle_propa} is $\gt_n = 2^{-n}$, and the matrix entries of the free propagator in position basis are given by Eq.~\eqref{eq_propa_circle_exact_finite}. The physical positions, according to Eq.~\eqref{eq_physical_even}, are the $N_n$ points identified on the circle by the discretization. This sets up a very convenient numerical approach to study the propagation also in presence of a potential. 

\begin{figure*}
\centering
\begin{subfigure}{0.33\textwidth}  
\includegraphics[scale=0.66]{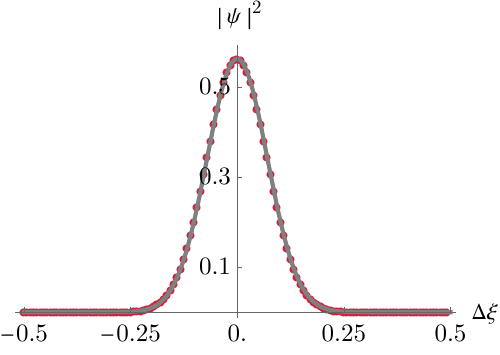}
\caption{}
\end{subfigure}\hfill
\begin{subfigure}{0.33\textwidth}  
\includegraphics[scale=0.66]{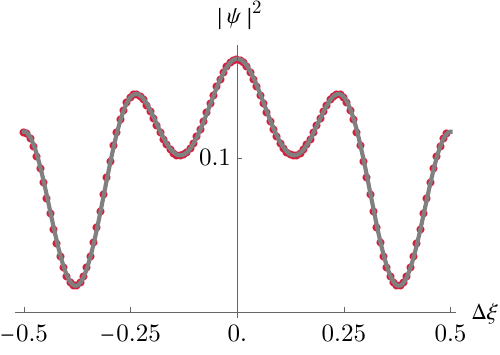}
\caption{}
\end{subfigure}\hfill
\begin{subfigure}{0.33\textwidth}  
\includegraphics[scale=0.66]{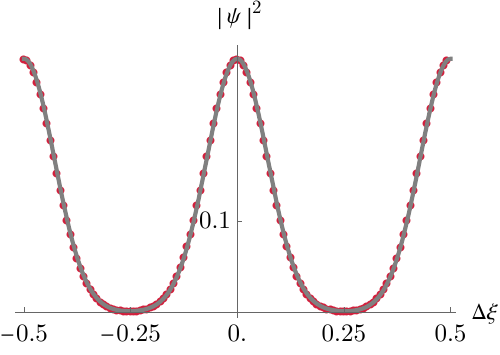}
\caption{}
\end{subfigure}
\caption{Plots of the probability density associated to free wave packets as function of the adimensional displacement for various simulation times $\gt$. The grey solid line is the analytical prediction Eq.~\eqref{e_time_evo_free_wave_packet_circle}, while the red dots are given by the numerical method exploiting Eq.~\eqref{e_path_integral_circle_propa}. The two match perfectly. The initial wave packet has parameters: $x_0 = 0$, $\kappa_0 =0$, $L\sigma =10$. The simulation has $n=7$, $N_n = 128$ points. The specific simulation times are: $\gt= 0$ (a), $\gt =1/4$ (b) and $\gt = 1/2$ (c).}
\label{f_time_evo_packet}
\end{figure*}

First, we benchmark this numerical approach with the analytical results in absence of potential, Eq.~\eqref{e_time_evo_free_wave_packet_circle}. In Fig.~\ref{f_time_evo_packet}, we show that the matching is faithful even for a reasonably small number of qubits, $n=7$ in this case. This is a nontrivial independent check of the validity of Eq.~\eqref{eq_propa_circle_exact_finite}.

\begin{figure*}
    \centering
    \begin{subfigure}{0.5\textwidth}  
\includegraphics[scale=0.6]{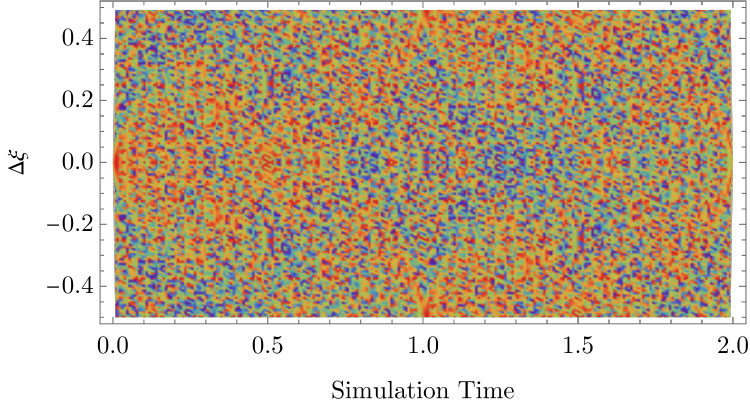}
\caption{}
\end{subfigure}\hfill
\begin{subfigure}{0.5\textwidth}  
\includegraphics[scale=0.6]{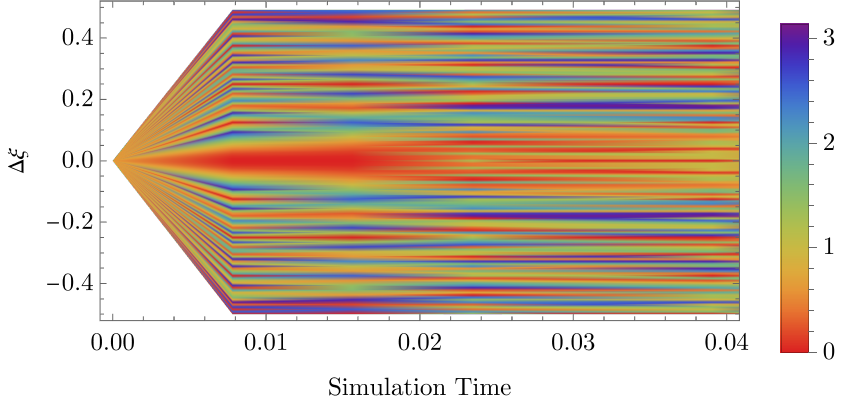}
\caption{}
\end{subfigure}
    \caption{Plot of the free propagator phase Eq.~\eqref{eq_propa_circle_exact_finite}, in absolute value, as function of the simulation time $\gt$ (horizontal axis) and adimensional displacement (vertical axis). The simulation has $n=7$, $N_n = 128$ points and the phase is simulated for multiples of the simulation time $\gt_n = 2^{-n}$. (a) presents the whole period, (b) only the early times $\gt \leq 0.04$. This plot can be seen as the phase of the wave function associated to a plane wave, for which the propagator provides the time evolution, retrieved in the limit of vanishing variance of the wave packet Eq.~\eqref{e_wave_packet_circle}. The minimum of phase (modulo $2\pi$) is red, while the maximum is in blue.}
    \label{f_phase_free_propagator}
\end{figure*}

\begin{figure*}
\centering
\begin{subfigure}{0.5\textwidth}  
\includegraphics[scale=0.66]{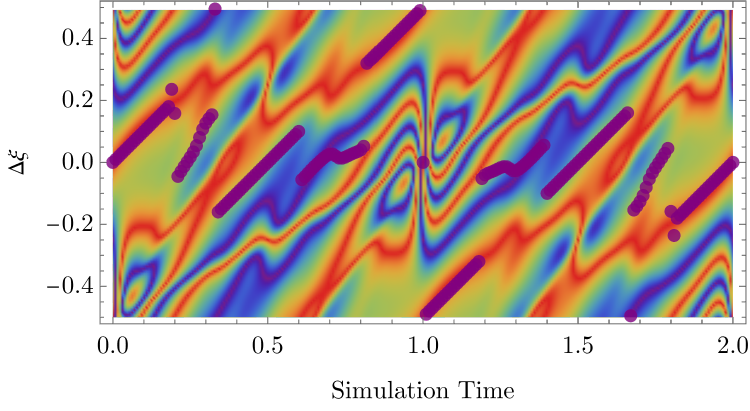}
\caption{}
\end{subfigure}\hfill
\begin{subfigure}{0.5\textwidth}  
\includegraphics[scale=0.66]{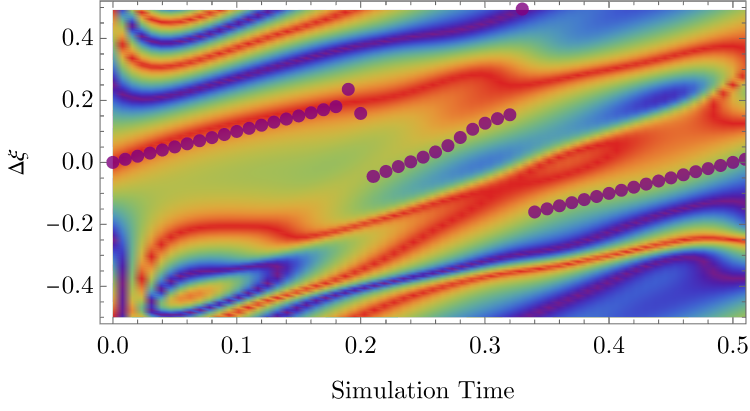}
\caption{}
\end{subfigure}
\caption{Plot of the absolute value of the time-evolved free wave packet phase, obtained from the numerical method, over the trajectory given by maximum of probability density Eq.~\eqref{e_time_evo_free_wave_packet_circle} (violet dots). The specific parameters are that of a packet moving with peak momentum $\kappa_0 =2\pi/L$ and is well-localized in position $L\sigma =10$ at $x_0 = 0$. The number of points of the simulation is $n=7$, $N_n = 128$. This plot is presented for the entire period (a) and quarter of period (b). The color gradient is the same as Fig.~\ref{f_phase_free_propagator}. At very early simulation times, the maximum of probability follows closely the minimum of wave function phase. As time goes on, wave packet delocalization and finite $L$ effects deviate the trajectory from the phase minimum.}
\label{f_maximum_proba_free}
\end{figure*}

It is well known that in the semi-classical limit $\hbar\to 0$, the maximum of the probability density follows the minimum of the propagator's phase. In terms of the simulation time $\gt$, this limit corresponds to small simulation times. We found most convenient not to study directly the phase of the free propagator, even though it is possible to glimpse at some hint of the semi-classical limit at early times, see Fig.~\ref{f_phase_free_propagator}, due to the simple fact that the free propagator assumes that the initial state is completely localized in position, and therefore every momentum contributes uniformly. Instead, simulating wave packets with small position variance, i.e. $a = \sigma L \gg 1$, provides a very convenient smearing, allowing us to emphasize this phenomenon, see Fig.~\ref{f_maximum_proba_free}.

\begin{figure*}
\centering
\begin{subfigure}{0.33\textwidth}  
\includegraphics[scale=0.66]{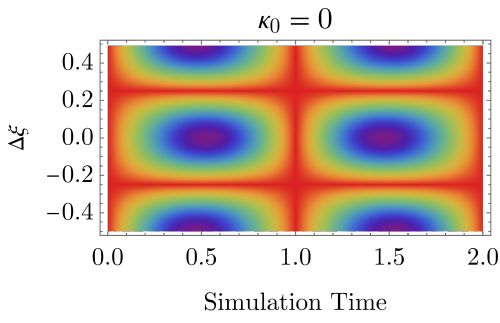}
\caption{}
\end{subfigure}\hfill
\begin{subfigure}{0.33\textwidth}  
\includegraphics[scale=0.66]{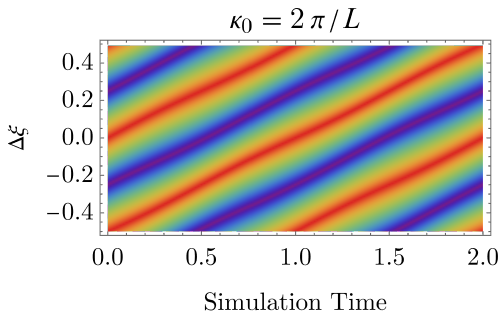}
\caption{}
\end{subfigure}\hfill
\begin{subfigure}{0.33\textwidth}  
\includegraphics[scale=0.66]{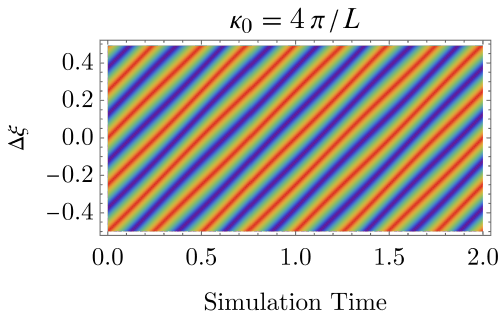}
\caption{}
\end{subfigure}
\caption{Plot of the absolute value of the time-evolved free wave packet phase, obtained from the numerical method, for a sequence of packets well-localized in momentum space $L\sigma =2.5$, $x_0 = 0$, varying the peak momentum $\kappa_0 = 0$ (a), $\kappa_0 = 2\pi/L$ (b) and $\kappa_0 = 4\pi/L$ (c). The number of simulation points is $n=7$, $N_n = 128$. The color gradient is the same as Fig.~\ref{f_phase_free_propagator}. From (b), we see a diagonal minimum of phase traveling to the right (upwards) and making $M_0=1$ windings around the circle over the period $\gt =2$. The diagonal line starting below at $\Delta \xi =-1/2$ is not a true phase minimum, because it is shifted by $2\pi$. In (c), the same diagonal line now makes $M_0=2$ windings around the circle over the period $\gt =2$. (b) and (c) justify the heuristic particle description Eq.~\eqref{e_particle_interpretation} that led to Eq.~\eqref{eq_propa_circle_exact_finite}. It is interesting to see that (a) presents no straight line corresponding to $M_0=0$, but it can be checked that the trajectory of the maximum is the constant line $\Delta\xi(\gt) =0$.}
\label{f_particle_trajectories}
\end{figure*}

On the other hand, packets with very large position variance $a\ll 1$ are useful in simulating the dynamics of packets peaked around some $\kappa_0$. If we look at the minimum of the phase for such free time-evolved wave function, it is possible to identify the semi-classical trajectory $x(t) = x_0 + v_0 t$ or alternatively $\Delta \xi(\gt) = M_0 \gt /2$, where $\kappa_0 = 2\pi M_0 / L$, $M_0 \in \integers$, see Fig.~\ref{f_particle_trajectories}. Very interesting is the case $M_0=0$, where no trajectory corresponding to $\Delta\xi(\gt) = 0$ can be found. We believe that this effect is due to positive $+\kappa$ and negative $-\kappa$ momenta to contribute equally to the wave packet. On the other hand, as long as $M_0 \neq 0$, the trajectory corresponding to $\Delta \xi(\gt) = M_0 \gt /2$ is the one corresponding to the phase minimum.

\begin{figure*}
\centering
\begin{subfigure}{0.5\textwidth}  
\includegraphics[scale=0.66]{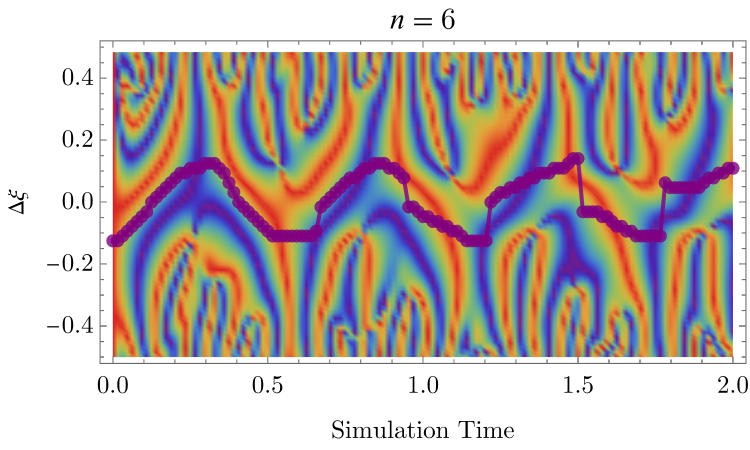}
\caption{}
\end{subfigure}\hfill
\begin{subfigure}{0.5\textwidth}  
\includegraphics[scale=0.66]{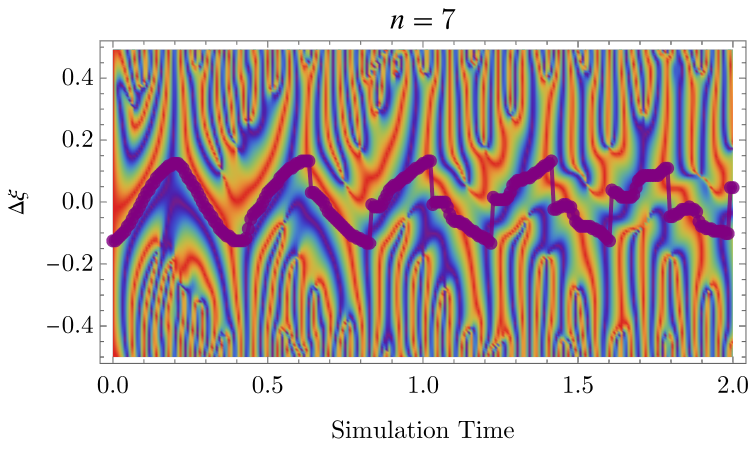}
\caption{}
\end{subfigure}
\caption{Plot of the absolute value of the time-evolved free wave packet phase, obtained from the numerical method, for a sequence of packets well-localized in position space $L\sigma =10$, $x_0 = -L/8$, $\kappa_0 = 0$,
for a harmonic potential Eq.~\eqref{f_harmonic} of $\eta_n = 3/4$. The violet dots represent the trajectory of the probability density maximum, also found numerically. Fixed $\eta_n$, we vary the number of qubits of discretization: $n=6$ (a), $n=7$ (b). The color gradient is the same as in Fig.~\ref{f_phase_free_propagator}. In both cases, the trajectory of the probability maximum is oscillating, as it is the case on the continuous line for the ground state wave function of the quadratic potential. It can also be seen that the lines corresponding to the minimum of the phase, modulo the smearing of the wave packet, tend to follow the trajectory of the probability maximum.}
\label{f_harmonic}
\end{figure*}

\begin{figure*}
\centering
\begin{subfigure}{0.5\textwidth}  
\includegraphics[scale=0.66]{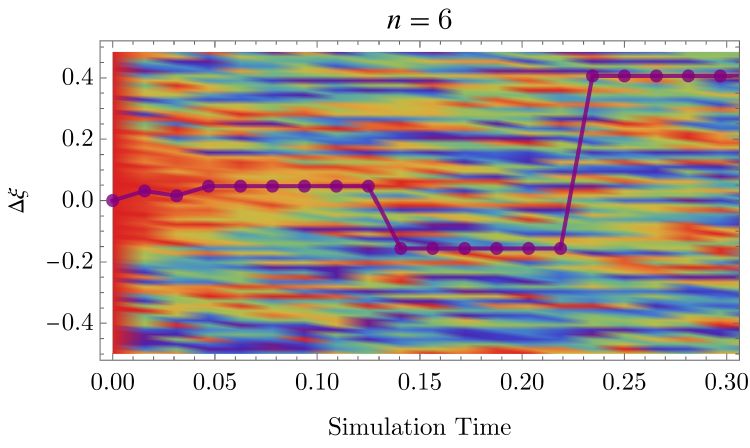}
\caption{}
\end{subfigure}\hfill
\begin{subfigure}{0.5\textwidth}  
\includegraphics[scale=0.66]{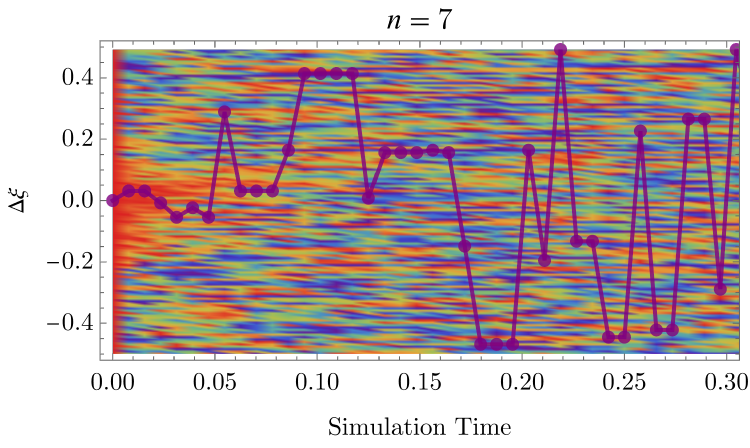}
\caption{}
\end{subfigure}
\caption{Plot of the absolute value of the time-evolved free wave packet phase for early simulation times $\gt \leq 0.3$, obtained from the numerical method, for a packet well-localized in position space $L\sigma =10$, $x_0 = 0$, $\kappa_0 = 0$,
for two realizations of a uniform random potential in in the range constrained by $\eta_n = 3/4$. The violet dots represent the trajectory of the probability density maximum, also found numerically. Fixed $\eta_n$, we vary the number of qubits of discretization. $n=6$ (a), $n=7$ (b). The color gradient is the same as Fig.~\ref{f_phase_free_propagator}. It is possible to see, at very early times $\gt\leq 0.05$, that the trajectory of the time evolved wave packet in the random potential follows the minimum of the propagator phase, smeared by the wave packet.}
\label{f_random}
\end{figure*}

With the numerical method presented, we computed the maximum of the time evolved wave function as a function of the simulated time, for two interesting potentials: the periodic one Eq.~\eqref{e_harmonic_potential}, which becomes the quadratic in the line limit and a uniformly distributed random one. In order to highlight the nontrivial dynamics given by the potential, we chose wave functions Eq.~\eqref{e_wave_packet_circle} with $\kappa_0 = 0$, i.e. with zero mean kinetic energy. These simulations show how delicate is the link between $\gt_n$ and the physical variables. Indeed, fixing the number of points (or qubits) fixes $\gt_n$, which in turn gives a constraint on the physical quantities $\hbar$, $m$, $L$ and $t$. Since in the path integral we need not only $\gt$ to be small, but also the product $\eta_n = (t_n/\hbar)\operatorname{max}\abs{V(x)}$, in order for the Trotter expansion not to be dominated by errors at large simulation times. Therefore $n$ not only fixes the smallest time scale, but also the maximum value of the potential. We found that fixing $\eta_n = 3/4$ allows us to obtain physically meaningful results. In Fig.~\ref{f_harmonic} we present the case of the harmonic potential Eq.~\eqref{e_harmonic_potential}, varying $n$. We noticed that increasing $n$ accentuates the short-time behavior expected from the dynamics of the quadratic potential: that of a maximum of probability density which is undergoing periodic motion. Qualitatively, the minimum of the action follows the maximum trajectory at early simulation times. This feature is also present when simulating a random potential taken from a uniform distribution constrained to satisfy $\eta_n = 3/4$, see Fig.~\ref{f_random}.

\section{Conclusions}\label{s_conclusions}

Motivated by results in digital quantum simulation, in this paper we analyzed the readout phase of a quantum algorithm simulating one-particle dynamics in one dimension. It turned out that the main ingredient for the analysis was the free propagator, which presents a different structure than the corresponding amplitude on a line. 

Indeed, we showed that the known case of a line was not a good starting point for the derivation of the free propagator on a circle, since it would entail reversing the limiting procedure from infinite $L$ to finite $L$. One of the main reasons is that on a line the free propagator has no periodicity, while it is always periodic as long as $L$ is finite. Another strategy was to find the free propagator on a circle divided into $N$ equidistant points. This setup is naturally implemented by the quantum algorithm of Sec.~\ref{s_rev_algo}.

When $N$ is finite, we showed that for rational simulation times $\gt = \frac{\hbar}{m} \frac{2\pi}{L^2} t = p/q$, the set of physical displacements, which are the possible outcomes of the free propagation of a localized particle, is finite. For $p=1$, we provided an exact formula for the free propagator, in which all physical positions are uniform in probability, and the phase is given by Eq.~\eqref{f_phase_free_propagator}, which is a continuous function of the squared displacement between the initial and final position, and inversly proportional to the simulation time, very similar in looks to that on a line. 

When the limit of infinite $N$ is considered, we retrieve the propagator on the continuum circle. In order to make sense of the $\sqrt{N}$ divergence, we needed to introduce a distribution $\sigma$, which is related to the Dirac $\delta$. On a circle, when the simulation time is infinitesimal, the set of allowed points becomes the whole circle, and it can be shown that the free propagator becomes that on the line. This is consistent with the fact that these infinitesimal simulation times are the only ones which survive the line limit $L\to\infty$, whatever the physical time $t$.

The knowledge of the free propagator with finite $N$ and $L$ is not only an analytical result, but sets up a very convenient numerical method to simulate the evolution of periodic wave functions, in presence or absence of potentials, see Sec.~\ref{s_numerical}. We confirmed that the numerical method in absence of potential gives the same results as the analytic predictions for free periodic wave packets. We showed that at very early simulation times, the trajectory given by the probability density maximum seems to follow the minimum of the -- smeared -- propagator phase, as one would expect in the semi-classical limit of quantum mechanics, in the free case and in presence of an harmonic and random potentials. 

In light of these results, it would be interesting to benchmark the quantum algorithm of one-particle dynamics with the current quantum computation platform and simulators. We think that this analysis provides solid grounds for the study of interactions between many bodies, especially in finite continuous and discrete geometries. It would be also interesting to address the problem of one-particle propagation taking into account relativistic effects.

\begin{acknowledgments}
The author thanks Giuseppe Mussardo for insightful discussions during the conceptualization and realization of the work. He also thanks Andrea Trombettoni and Gioele Zambrotti for discussions and feedback on the manuscript.
\end{acknowledgments}

\bibliography{readout_biblio}

\clearpage
\appendix
\section{Uncertainty Relations in Different Geometries: Line, Segment, Circle and Lattice}\label{a_uncertainty} 

In this Appendix, we address the problem of giving a proper meaning to a momentum operator in geometries which are not the continuous (infinite) line, for which the theory of one-particle dynamics is well known. We investigate two geometries, related to the line by two different limits: the circle and the lattice. On the circle, positions are continuous while momenta are discrete. On the lattice, the converse. 

The simplest definition of ``momentum operator'' given an Hilbert space of physical wave functions on some geometry, for which we have a notion of position, involves completeness as a basis, and satisfaction of the uncertainty relations. Let us begin by reviewing the case of the line.

\subsection{Line}

We would like to investigate the nature of the momentum operator $K$, starting from this setup. Suppose we perform an ideal experiment where the only means of detecting a particle is through ideal detectors. If at a certain time all detectors are measuring simultaneously, we will obtain an outcome of a position measurement $x$. Naturally, this setup is not realistic, since to sample the line, both uncountable many point-like detectors and the possibility of accessing the whole line are required. The limit of the continuous line is twofold then: first, for a finite interval $L$, the number $N$ of detectors is let to infinity. This retrieves the continuum limit of a segment -- or a circle, depending on the physical wave functions considered. If subsequently $L$ is sent to infinity, the segment becomes the line. These two limits are not interchangeable. If the particle were classical, measurements of position at different instances of time allow us to find also the velocity, simultaneously with the position. As well known, this is not so in the case of quantum mechanics, since simultaneous measurements of position and momentum are not independent. 

From the ideal experiment, similarly to Sec.~\ref{s_rev_oneparticle}, the completeness of position measurements -- we know that the particle can be found somewhere on the line -- sets up a position basis of the one-particle Hilbert space, denoted by $\ket{x}$, eigenvectors of the position operator $X$ with eigenvalue $x\in \reals$. As such, $X$ has continuous unbounded spectrum and its eigenstates are $\delta$-normalized:
\begin{equation}
    \braket{x'|x} \defi \delta(x'-x).
\end{equation}

It is also true that complete knowledge of particle momentum as a function of time, neglecting all information about position, defines the particle state. As such, momentum is just another basis for the one-particle Hilbert space. For such basis, labeled $\ket{k}$, it is found according to these properties: 
\begin{enumerate}
    \item It has states which are orthonormal.
    \item Any state $\ket{x}$ can be written as a linear combination of the $\ket{k}$.
    \item It diagonalizes an operator $K$ with eigenvalue $k$. 
    \item The operator $K$ can not be diagonalized simultaneously with $X$; in particular on one-particle physical states $\ket{\Psi}$:
    \begin{equation}
        \braket{\Psi|\comm{X}{K}|\Psi} = i.
    \end{equation}
\end{enumerate}

The quantum Fourier Transform, already outlined in Sec.~\ref{s_rev_oneparticle} for the circle, allows us to satisfy the first three properties. We build the states $\ket{k}$ out of uniform superpositions of $\ket{x}$:
\begin{equation}
    \ket{k} = \intmpinf{x} \frac{e^{ i k x}}{\sqrt{2\pi}} \ket{x}, \quad \bra{k} = \intmpinf{x} \frac{e^{-i k x}}{\sqrt{2\pi}} \bra{x}. 
\end{equation}
The first property, orthonormalization, can be readily checked
\begin{equation*}
    \braket{k'|k} = \intmpinf{x'} \intmpinf{x} \frac{e^{i(kx-k'x')}}{2\pi} \braket{x'|x} = \inv{2\pi} \intmpinf{x} e^{i (k-k')x} = \delta(k-k').
\end{equation*}

On the line, the space of physical states is those for which the (position) wave function $\psi(x)$ is square-integrable: $L^2(\reals)$. At the same time, since by definition also $\ket{k}$ is a basis, an analogous momentum wave function $\Psi(k)$ also exists
\begin{equation}
    \ket{\Psi} = \intmpinf{x} \psi(x)\ket{x} = \int_{D_K} \dd k\, \Psi(k) \ket{k}.
\end{equation}
Here $D_K$ is the domain in which $k$ takes values, i.e. the support of the operator $K$ for which $\ket{k}$ is diagonal with eigenvalue $k$. It is the content of the third point to determine $D_K$. 

As well known, the range of $k$'s is $\reals$. Indeed, the first way we can check it is by simple use of the definition
\begin{equation*}
    \int_{D_K} \dd k\, \Psi(k) \ket{k} = 
    \int_{D_K} \dd k\, \Psi(k) \intmpinf{x} \frac{e^{ikx}}{\sqrt{2\pi}} \ket{x} \coinc \intmpinf{x} \psi(x)\ket{x}.
\end{equation*}
Moreover, $\psi(x)$ is a square-integrable function, admitting a regular Fourier transform in terms of $\kappa \in \reals$:
\begin{equation*}
    \intmpinf{x} \psi(x)\ket{x} \coinc \intmpinf{x} \intmpinf{\kappa} \Tilde{\Psi}(\kappa) \frac{e^{i\kappa x}}{\sqrt{2\pi}} \ket{x} = \int_{D_K} \dd k\, \Psi(k) \intmpinf{x} \frac{e^{ikx}}{\sqrt{2\pi}} \ket{x}.
\end{equation*}
We would like to identify $k \coinc \kappa$, $D_K \coinc \reals$ and $\Psi(k) \coinc \Tilde{\Psi}(\kappa)$. To do so, we can exploit the fact that the equality holds for any choice of $\psi(x)$. For $\psi(x) = \delta(x-x_0)$ we obtain $\Tilde{\Psi}(\kappa) = e^{-i \kappa x_0}/\sqrt{2\pi}$, yielding
\begin{equation*}
    \ket{x_0} \coinc \int_{D_K} \dd k\, \Psi(k) \intmpinf{x} \frac{e^{ikx}}{\sqrt{2\pi}} \ket{x}.
\end{equation*}
The only way we can obtain the equality, for all $x_0 \in \reals$, is to consider $\Psi(k) = e^{-i k x_0}/\sqrt{2\pi}$ and $D_K = \reals$. Thus $D_K = \reals$, which in turn implies $k \coinc \kappa$, since they are simply two ways of labelling the same quantity. From the last identification, back into the definition of a generic state, under the reasonable assumption of convergence of the wave function,
\begin{equation*}
    \ket{\Psi} = \intmpinf{k} \Psi(k) \ket{k} = \intmpinf{x} \sbr{\inv{\sqrt{2\pi}} \intmpinf{k} \Psi(k) e^{ikx} } \ket{x},
\end{equation*}
it is immediate to identify $\Psi(k)$ with the ordinary Fourier Transform of $\psi(x)$, for any square integrable $\psi(x)$. Most importantly, from this reasoning we can verify the second, i.e. that position states are uniform superpositions of $\ket{k}$:
\begin{equation}
    \ket{x} = \intmpinf{k} \frac{e^{-ikx}}{\sqrt{2\pi}} \ket{k}.
\end{equation}
Compared to Sec.~\ref{s_rev_oneparticle}, the matrix element of the quantum Fourier Transform are
\begin{equation}
    \braket{x|k} = \frac{e^{ikx}}{\sqrt{2\pi}}, \quad \braket{k|x} = \frac{e^{-ikx}}{\sqrt{2\pi}}.
\end{equation}

Since the two basis are complete, we can write two resolutions of the identity:
\begin{equation}
    \intmpinf{x}\pro{x} = \id = \intmpinf{k} \pro{k}.
\end{equation}
Furthermore, the operator $K$ is the one diagonalized by $\ket{k}$ and the spectrum is continuous and unbounded:
\begin{equation}
    K \ket{k} \defi k \ket{k}, \quad \forall k \in\reals
\end{equation}

On the line, the quantum Fourier Transform allows to retrieve the well-known differential form for $K$ with a one-line computation:
\begin{equation}
    K\ket{k} = k \intmpinf{x} \frac{e^{ikx}}{\sqrt{2\pi}} \ket{x} = \intmpinf{x} (-i \dv_x) \frac{e^{ikx}}{\sqrt{2\pi}} \ket{x} := (-i \dv_{X}) \intmpinf{x}\frac{e^{ik X}}{\sqrt{2\pi}} \ket{x} \follows K =: -i \dv_X.
\end{equation}
Notice that the last one is an operator identity which formally defines $\dv_X$, not vice versa. Along similar lines one can find
\begin{equation}
    X\ket{x} = x\intmpinf{k}\frac{e^{-ikx}}{\sqrt{2\pi}} \ket{k} = \intmpinf{k}(i \dv_k) \frac{e^{-ikx}}{\sqrt{2\pi}} \ket{k} := (i\dv_K) \intmpinf{k} \frac{e^{-ikx}}{\sqrt{2\pi}} \ket{k}\follows X=: i \dv_K.
\end{equation}

The last point, concerning the uncertainty relation, can be checked by using all the results previously obtained. We recall that an operator is defined by its matrix elements and therefore specifying all the entries in a particular basis allows us to find the operator. We check it in the basis of $K$
\begin{equation*}
    \begin{split}
        \braket{k| \comm{X}{K} |k'} &= \intmpinf{x}\sbr{\braket{k|X|x}\braket{x|K|k'} -  \braket{k|K|x}\braket{x|X|k'}}\\
        &= \intmpinf{x} x (k'-k) \frac{e^{i(k'-k)x}}{2\pi}\\
        &= -i \intmpinf{x} x \dv_x \frac{e^{i(k'-k)x}}{2\pi} \underset{\text{equivalently}}{=} -i (k'-k) \dv_{(k'-k)} \delta(k'-k)\\
        & \underset{\text{I.B.P.}}{=} -i\sbr{x\frac{e^{i(k'-k)x}}{2\pi}}_{-\infty}^\infty + i \delta(k'-k).
    \end{split}
\end{equation*}
We can also find $\comm{X}{K}$ in the $X$-basis but the computation is not so dissimilar. This will not be the case on the circle or the lattice, for momenta and position have different spectra. 
Notice that the vanishing of the boundary term is essential to retrieve the correct commutation relations for momentum and position. Physical states on the line ensure this. Thus on these states:
\begin{equation*}
    \braket{\Psi|\comm{X}{K}|\Psi} = \intmpinf{k}\intmpinf{k'} \Psi(k)\Psi^*(k') \braket{k| \comm{X}{K} |k'} = i.
\end{equation*}

\subsection{Segment}

Let us consider a physical particle, constrained to live on a finite segment of length $L$. We can imagine the size $L$ to be the extension of the laboratory in which we are conducting the ideal experiment. Any measurement outside $L$ of position is always happening with probability zero. This means that, if the laboratory has coordinates $x\in\sbr{-L/2, L/2}$, that any physical state will have wave function
\begin{equation}
    \ket{\Psi} = \int_{-L/2}^{L/2} \dd x\, \psi(x) \ket{x}.
\end{equation}
In comparison to the previous case, the position operator $X$ has a continuous and bounded spectrum $\sbr{-L/2, L/2}$. Therefore conservation of probability yields
\begin{equation}
    \int_{-L/2}^{L/2} \dd x\, \norm{\psi(x)} = 1.
\end{equation}
We may imagine the walls of the laboratory to constitute an infinitely high potential barrier, since no particle can escape it. As such the probability density should vanish:
\begin{equation}
    \psi(-L/2) = 0 =\psi(+L/2).
\end{equation} 
On the other hand, the spectrum of $X$ being continuous implies $\braket{x|x'} = \delta(x-x')$
for any $x$, $x'$, inside or outside of the laboratory. The constraint on position is reflected in the resolution of the identity
\begin{equation}
    \id = \int_{-L/2}^{L/2} \dd x\, \pro{x}.
\end{equation}
This operator is different from that of the previous section; in fact, different Hilbert spaces are describing the particle: for the line it is that of square-integrable wave functions, for the particle on the segment it is that of wave functions which are nonzero only in the laboratory and vanish at the boundary.

This means that there are some position eigenstates $\ket{\chi}$ which are guaranteed not to contribute to any physical state. These positions are those outside the laboratory and are such that
\begin{equation}
    \braket{\chi|\Psi} \coinc 0, \quad \forall \chi \notin \sbr{-L/2, L/2}.
\end{equation}
From now on, this prohibited states will be called $\ket{\chi}$, while allowed states $\ket{x}$.

For a finite segment, denote $\ket{\kappa}$ the momentum eigenstates. As in the previous section and Sec.~\ref{s_rev_oneparticle}, we define them through a quantum Fourier Transform
\begin{equation}
    \ket{\kappa} = \int_{-L/2}^{L/2} \dd x\, \frac{e^{i\kappa x}}{\sqrt{L}} \ket{x}.
\end{equation}
As before, we let momenta be a basis for the physical Hilbert space, i.e. we write
\begin{equation}
    \ket{\Psi} = \int_{-L/2}^{L/2} \dd x\, \psi(x) \ket{x} = \sum_{\kappa\in\momentumlattice{L}} \Psi_\kappa \ket{\kappa}, 
\end{equation}
where the coefficient in the basis are nothing more than the Fourier transform of the wave function in position space
\begin{equation}
    \Psi_\kappa = \int_{-L/2}^{L/2} \dd x\, 
    \frac{e^{-i \kappa x}}{\sqrt{L}} \psi(x), \quad \psi(x) = \sum_{\kappa \in \momentumlattice{L}} \frac{e^{i \kappa x}}{\sqrt{L}} \Psi_\kappa.
\end{equation}
The constraint on the position wave function translates to
\begin{equation}
    \sum_{\kappa \in \momentumlattice{L}} \frac{e^{i \kappa L/2}}{\sqrt{L}} \Psi_\kappa =0.
\end{equation}

Therefore the states $\ket{\kappa}$ are eigenstates of a \textit{discrete} momentum operator $K$ with eigenvalues $ \kappa \in \momentumlattice{L}$. Since the operator $K$ has discrete spectrum, we have
\begin{equation}
    \braket{\kappa'|\kappa} = \krdel{\kappa'}{\kappa}, \quad \forall\, \kappa', \, \kappa \in \momentumlattice{L}.
\end{equation}
The projections onto states of physical position are
\begin{equation}
    \braket{x|\kappa} = \frac{e^{i \kappa x}}{\sqrt{L}}, \quad \braket{\kappa|x} = \frac{e^{-i \kappa x}}{\sqrt{L}}.
\end{equation}

Naturally, the $\kappa$ are associated to momenta as long as we give a physical interpretation for them. Therefore, we must ensure the uncertainty relation only on physical states. Let us compute the matrix elements of $\comm{X}{K}$ in the basis of position and momenta. 

For the projections onto momenta states two equivalent methods can be used. The first one makes use of integration by parts, which is perfectly legitimate since position states are continuous. The second one exploits the analytic continuation to continuous momenta. They both yield the same result. The first one gives
\begin{equation*}
\begin{split}
    \braket{\kappa| \comm{X}{K}|\kappa'} = \int_{-L/2}^{L/2} \dd x\, x (\kappa-\kappa') \frac{e^{-i (\kappa - \kappa') x}}{L} &\underset{\text{I.B.P.}}{=} -i \sbr{x\frac{e^{-i (\kappa - \kappa') x}}{L}}_{-L/2}^{L/2} + i \int_{-L/2}^{L/2} \dd x\, \frac{e^{-i (\kappa - \kappa') x}}{L} \\
    &= -i e^{-i(\kappa-\kappa')L/2} + i \krdel{\kappa}{\kappa'}.
    \end{split}
\end{equation*}
In the last line we used the fact that $\kappa-\kappa'$ is an integer multiple of $2\pi/L$. In the second case, we have 
\begin{equation*}
    \begin{split}
    \braket{\kappa| \comm{X}{K}|\kappa'} = \int_{-L/2}^{L/2} \dd x\, x (\kappa'-\kappa) &\frac{e^{-i (\kappa - \kappa') x}}{L} \underset{\text{I.B.P.}}{=}i (\kappa'-\kappa)\dv_{(\kappa-\kappa')} \int_{-L/2}^{L/2} \dd x\, \frac{e^{-i (\kappa - \kappa') x}}{L}\\
    &=i (\kappa'-\kappa) \dv_{(\kappa-\kappa')}\frac{e^{-i (\kappa - \kappa') L/2}-e^{i (\kappa - \kappa') L/2}}{-i L(\kappa-\kappa')}\\
    &= i (\kappa'-\kappa) \br{\frac{e^{-i (\kappa - \kappa') L/2}-e^{i (\kappa - \kappa') L/2}}{i L (\kappa-\kappa')^2} + \frac{e^{-i (\kappa - \kappa') L/2}+e^{-i (\kappa - \kappa') L/2}}{2(\kappa-\kappa')}}\\
    &= -i e^{-i (\kappa - \kappa') L/2} + i \krdel{\kappa}{\kappa'},
    \end{split}
\end{equation*}
where in the last line we used the fact that the sum is $-i e^{-i(\kappa-\kappa')L/2}$ for all $\kappa-\kappa'$ multiple integers of $2\pi/L$, while we have used a small coupling expansion for $\kappa = \kappa'$. 

Both methods lead to the same result: in the momenta basis it is off-diagonal and each entry is $-i e^{-i (\kappa - \kappa') L/2}$:
\begin{equation}
    \comm{X}{K} = -i \sum_{\kappa}\sum_{\kappa \neq \kappa'} e^{-i(\kappa-\kappa')L/2}\ket{\kappa}\bra{\kappa'}.
\end{equation}
One would naively be led to suspicion, since this is not so in the continuous case, since in that case $\comm{X}{K} = i$, and therefore the matrix entries are always diagonal. This emphasizes the importance of considering the right states onto which we compute $\comm{X}{K}$. For it indeed on physical states, we have
\begin{equation}
\begin{split}
    \braket{\Psi|\comm{X}{K}|\Psi} &= -i \sum_{\kappa}\sum_{\kappa \neq \kappa'} \Psi_\kappa^* \Psi_{\kappa'}e^{-i(\kappa-\kappa')L/2} 
    \\
    &= -i \sbr{\sum_\kappa \Psi_\kappa^* e^{-i \kappa L/2}\sum_{\kappa'}\Psi_{\kappa'}e^{i \kappa' L/2} - \sum_\kappa \norm{\Psi_\kappa}} \\
    &= -i L \norm{\sum_{\kappa}\Psi_{\kappa}\frac{e^{i \kappa L/2}}{\sqrt{L}}} + i \underset{\psi(\pm L/2) = 0}{=} i.
    \end{split}
\end{equation}

As a further check, let us retrieve the same result using position as basis. We have
\begin{equation*}
    \begin{split}
    \braket{x|\comm{X}{K}|x'} &= \sum_{\kappa\in\momentumlattice{L}}\kappa(x-x') \frac{e^{i \kappa (x-x')}}{L} = -i (x-x') \dv_{(x-x')} \sum_{J\in\integers} \delta(x-x' + JL) = -i\sbr{ \xi \dv_{\xi}\sum_{J\in\integers} \delta(\xi+JL)}_{\xi=x-x'}.
    \end{split}
\end{equation*}
Therefore in position space the commutator assumes the form
\begin{equation*}
    \begin{split}
        \comm{X}{K} &= \int_{-L/2}^{L/2} \dd x\, \int_{-L/2}^{L/2} \dd x'\, \cbr{-i\sbr{ \xi \dv_{\xi}\sum_{J\in\integers} \delta(\xi+JL)}_{\xi=x-x'}} \ket{x'}\bra{x}\\
        &\underset{\text{I.B.P} }{=}  \int_{-L/2}^{L/2} \dd x\, \cbr{i \br{x- \frac{L}{2}} \sbr{\delta \br{x- \frac{L}{2}}\ket{\frac{L}{2}}+\delta \br{x+ \frac{L}{2}}\ket{-\frac{L}{2}}}}\bra{x}+\\
        &\quad + \int_{-L/2}^{L/2} \dd x\, \cbr{-i \br{x+ \frac{L}{2}} \sbr{\delta \br{x- \frac{L}{2}}\ket{\frac{L}{2}}+\delta \br{x+ \frac{L}{2}}\ket{-\frac{L}{2}}}}\bra{x}+\\
        &\quad \int_{-L/2}^{L/2} \dd x\, \int_{x+L/2}^{x-L/2} \dd\xi\, \sbr{i \delta(\xi)\xi\dv_\xi}\ket{x-\xi}\bra{x} + \\
        &\quad +i \int_{-L/2}^{L/2} \dd x\, \pro{x}.
    \end{split}
\end{equation*}
Notice that the operator reduces to
\begin{equation}
\begin{split}
   \comm{X}{K}&= i \int_{-L/2}^{L/2} \dd x\,  \br{x- \frac{L}{2}} \delta\br{x+ \frac{L}{2}} \ket{-\frac{L}{2}}\bra{x} + \\
   &\quad -i \int_{-L/2}^{L/2} \dd x\,  \br{x+ \frac{L}{2}} \delta\br{x-\frac{L}{2}} \ket{+\frac{L}{2}}\bra{x} + \\
   & \quad + i \int_{-L/2}^{L/2} \dd x\, \pro{x},
   \end{split}
\end{equation}
because terms like $\xi\delta(\xi)$ always cancel in integration. As before, we compute the commutation relations on physical states
\begin{equation}
    \braket{\Psi|\comm{X}{K}|\Psi} = -i L \sbr{\norm{\psi(-L/2)} + \norm{\psi(L/2)}} + i = i.
\end{equation}
Thus $K$ is a legitimate momentum operator on the segment.

\subsection{Circle}

The structure of momenta and uncertainty relations of the particle on a circle of length $L$ is found along similar lines to the previous case, the segment of same length $L$. The only difference is in the Hilbert space of physical states: for the particle on the circle it is given by the set of periodic wave functions with period $L$. The formulas for the quantum Fourier Transform have already been given in Sec.~\ref{s_rev_oneparticle}. Here we show how the commutation relations for position and momenta on the circle can be obtained starting from those on a segment.

Let us consider a measurement of position for a particle on a circle. It would read a value in between $a$, the minimum value, and $a+L$, the maximum, where $a$ is the natural offset of the detector. In the case of the previous ideal experiment, $a= -L/2$. In absence of any coupling to external sources, which may be of the type of Aharanov-Bohm, it is impossible to distinguish the number of times the particle has circled around the line, the winding number, before measuring. This impossibility of detecting the winding number of the particle implies that the position operator is degenerate. Let us call $\Xi$ this new position operator on the circle and therefore it has eigenstates $\ket{\xi}$ with infinitely-degenerate eigenvalues
\begin{equation}
    \Xi \ket{\xi} = (\xi \operatorname{mod} L) \ket{\xi}.
\end{equation}

The resolution of identity would then consider all the possible outcomes of a position measurement. As long as we can not distinguish the winding number, the identity reads
\begin{equation}
    \id_a = \int_{a}^{a+L} \dd \xi \pro{\xi}. 
\end{equation}
It seems that this identity is depending on the particular value of $a$, but the $\ket{\xi}$ can only be distinguished by their eigenvalues of $\Xi$. And the eigenvalues are degenerate. So, there is no contradiction in considering an identity with two different $a$'s, since they would simply imply a different reference point of the ideal detector. As such, an identity which is insensitive to the particular choice of $a$ is
\begin{equation}
    \id \defi \int_{0}^L \frac{\dd a}{L} \id_a = \int_{0}^L \frac{\dd a}{L} \int_{a}^{a+L} \dd \xi \pro{\xi} \defi \intreg{L}{\xi} \pro{\xi}.
\end{equation}
In the last equation, we simply defined the integral over a period. 

As a consequence any physical state $\ket{\Phi}$ will have wave function
\begin{equation}
    \ket{\Phi} = \intreg{L}{x} \phi(\xi) \ket{\xi}, \quad \phi(\xi) = \phi(\xi+nL), \quad n\in\integers.
\end{equation}
As far as momentum is concerned, we also define these states whose projection onto position states is
\begin{equation}
    \braket{\xi|\kappa} \coinc \frac{e^{i \kappa (\xi\operatorname{mod}L)}}{\sqrt{L}}.
\end{equation}

To find a straightforward expression for these momentum states is not immediate. Define the ``comb'' -- i.e. periodic Dirac $\delta$-- of period $L$
\begin{equation}
    \gd(\xi) = \sum_{m\in\integers} \delta(\xi + m L).
\end{equation}
Let us consider a projector very similar to the identity
\begin{equation*}
    \intmpinf{\xi} \gd(\xi) \pro{\xi} = \sum_{m\in \integers} \int_{a + m L}^{a+ (m+1) L} \dd\xi\, \pro{\xi}\sim \sum_{m\in \integers} \id_{a+mL}.
\end{equation*}
Formally, this projector is an ``unregulated'' definition of the identity. And since the l.h.s does not depend explicitly from $a$, we must also have that the r.h.s does too. Because of this, we have
\begin{equation}
    \intmpinf{\xi} \gd(\xi) \pro{\xi} = \sum_{m\in \integers} \id.
\end{equation}
As a consequence, we define the unregulated momenta as those which satisfy
\begin{equation}
    \int_{-\infty}^\infty \gd(\xi) \braket{\xi|\kappa} \ket{\xi}   =\sum_{m\in\integers}  \int_{a + m L}^{a+ (m+1) L} \dd\xi\, \frac{e^{i \kappa (\xi\operatorname{mod}L)}}{\sqrt{L}} \ket{\xi} \coinc \sum_{\kappa \in \momentumlattice{L}}\ket{\kappa}.
\end{equation}
Again, the l.h.s. does not depend on the particular choice of $a$, and therefore the last term of the equation does too. Regularized momenta states are therefore defined by inserting the regulated identity:
\begin{equation}
    \ket{\kappa} \defi \intreg{L}{\xi} \frac{e^{i \kappa \xi}}{\sqrt{L}} \ket{\xi}, \quad \kappa \in \momentumlattice{L}.
\end{equation}

In order to find the expression of the commutation relations, let us recall that whenever we fix $a$, the commutation relation of position $\Xi$ and momenta $K_a$
\begin{equation}
    \braket{\kappa_a|\comm{X}{K_a}|\kappa'_a} = -i e^{i(\kappa-\kappa')a} + i \krdel{\kappa_a}{\kappa'_a}.
\end{equation}
where the subscript implies that we start measuring position from a fixed $a$. However, the momenta state we defined do not depend on $a$, therefore
\begin{equation}
    \braket{\kappa|\comm{X}{K}|\kappa'} = -i \int_{0}^L \frac{\dd a}{L} e^{i(\kappa-\kappa')a} + i  \int_{0}^L \frac{\dd a}{L}\krdel{\kappa}{\kappa'} = i \krdel{\kappa'}{\kappa},
\end{equation}
where the last equality holds because the first term cancels, since $(\kappa-\kappa')$ is integer multiple of $2\pi/L$. Then for physical states with periodic wave functions, the momenta defined by summing over all possible periods ensure that canonical commutations relations are preserved
\begin{equation}
    \braket{\Phi|\comm{\Xi}{K}|\Phi} = i.
\end{equation}

\subsection{Lattice}

Let us consider an experiment in which we place a set of infinitely many equidistanced detectors on a line. Each detector will read, upon the detection of particle, a position
\begin{equation}
    x_J = J \ell, \quad J\in\integers.
\end{equation}
with $\ell$ is the lattice spacing. 

Even though position is a continuous operator, the set of allowed states which we can measure through the detectors is discrete. They are labeled $\ket{J}$, different from $\ket{x_J}$ which are by definition continuous. The link between these states, for a line of equidistanced points is given by the equations
\begin{equation}
    \ket{J} \defi \int_{x_J \pm \half \ell} \dd x\, \frac{1}{\sqrt{\ell}} \ket{x}, \quad \ket{x_J} = \Lim{\ell}{0} \sqrt{\ell} \ket{J}, 
\end{equation}
where the limit is strictly true for expectation values of observables, i.e. when considering $\braket{J|\ldots|J}$. 

Indeed the states $\ket{x_J}$ and $\ket{J}$ differ for finite $\ell$, since measurements of position on $\ket{J}$ yield
\begin{equation}
    \braket{X}_J = x_J, \quad \braket{X^2}_J - \braket{X}_J^2 = \frac{\ell^2}{12},
\end{equation}
which are the classical expectation values for a random variable distributed in the interval $\sbr{x_J - \half \ell, x_J + \half \ell}$.

Indeed, these relation follow from the simple computations:
\begin{equation*}
    \braket{J|X|J} = \int_{x_J \pm \half \ell} \dd x'\, \int_{x_J \pm \half \ell} \dd x \, x \delta(x-x') \norm{\braket{x|J}} = \frac{\br{x_J + \half\ell}^2 - \br{x_J - \half\ell}^2}{2\ell} = x_J,
\end{equation*} 
\begin{equation*}
    \braket{J|X^2|J} = \int_{x_J \pm \half \ell} \dd x'\, \int_{x_J \pm \half \ell} \dd x \, x^2 \delta(x-x') \norm{\braket{x|J}} = \frac{\br{x_J + \half\ell}^3 - \br{x_J - \half\ell}^3}{3\ell} = x_J^2 + \frac{\ell^2}{12}.
\end{equation*}

Since the set of $\ket{J}$ is discrete, any sequence $\cbr{f_J}$ can be square-normalized and therefore can constitute a state of the Hilbert space. Naturally, it is only those sequences which vanish in the limit $J\to\infty$ which are samplings of square-integrable position wave functions on the line. Within experimental error $\ell$, a measurement of position at a certain time will correspond to one detector $J$ detecting the particle. Therefore the set $\ket{J}$ is a complete basis in this sense. 

Its conjugate basis, $\ket{K}$, is given by the quantum Fourier Transform
\begin{equation}
    \ket{K} = \sum_{J\in\integers} e^{i K x_J} \ket{J}
\end{equation}
and corresponds to a set of momenta which are continuous and bounded in an interval of length $2\pi/\ell$, which we choose to be $[-\pi/\ell, \pi/\ell)$. The momentum wave function are obtained by taking the appropriate Fourier Transform and are periodic with period $2\pi/\ell$.

It is immediate to realize that we can interpret the positions $J$ on the lattice as the momenta $\kappa$ on the circle of the previous subsection, and the momenta $K$ on the lattice as the positions $\xi$ on the circle. From the point of view of Hilbert spaces, the two problems are equivalent and therefore $J$ and $K$ are conjugate bases. The commutation relations between the operators $J$ and $K$ on the physical states follows straightforwardly from those derived on a circle.

\section{Technical Results}\label{a_technical}

In this Appendix we address the computation of Eq.~\eqref{eq_phase_0_q_even} in Sec.~\ref{s_propa_circle_finite_N} in the main text. Let us consider
\begin{equation}
   S_Q^p = \sum_{M=-2Q}^{2Q-1} e^{-i 2\pi p \frac{M^2}{4Q}} = 2\sum_{M=0}^{2Q-1} e^{-i 2\pi p \frac{M^2}{4Q}},
\end{equation}
which reduces to Eq.~\eqref{eq_phase_0_q_even} for $q = 2Q$ and $p=1$.\comment{ First of all, notice that it is difficult to apply a heuristic wave interference argument even in the simplest case $p=1$. In fact, one can immediately realize that at multiple cases have to be considered, just by looking at the intermediate value $M=Q$, for which the phase reads $\exp\cbr{-i \pi Q/2}$ and can attain four different values, varying $Q$.} The strategy we will pursue in this appendix is to consider $S_Q^p$ as sum of residues of some analytic function, and to use the fact that such sum is related to the residue at infinity. We are then supposing that  
\begin{equation}
    S_Q^p = \oint \frac{\dd z}{2\pi i} f(z),
\end{equation}
where $f(z)$ is a function with poles at $2Q$ distinguished points, with residue $2 \exp\cbr{-i \pi M^2 /(2Q)}$, for $M= \cbr{0, 1, \ldots, 2Q-1}$. The contour of integration encloses all points. Therefore, we can write
\begin{equation}
    S_Q^p = - \operatorname{Res}_\infty f(z).
\end{equation}
We have reduced the problem to the computation of a single residue at infinity.

There is more than one such function. The one we found most convenient is
\begin{equation}
    f(z) = e^{- i \pi p \frac{z^2}{2Q}} \sum_{M=0}^{2Q-1} \inv{z-M} =\comment{ e^{- i \pi p \frac{z^2}{2Q}} \sbr{H(z) - H(z-2Q)} =} e^{- i \pi p \frac{z^2}{2Q}} \sbr{\psi(z) - \psi(z-2Q) + \br{\inv{z}-\inv{z-2Q}}},
\end{equation}
which has simple poles at $z=M$. Here $\psi(z)$ is the digamma function \cite{NIST}
\begin{equation*}
    \psi(z) = \dv_{z} \ln \ega{z} = - \gamma + \sum_{n=1}^\infty \br{\inv{n} - \inv{z+n}}, \quad \psi(z+1) = \psi(z) + \inv{z}, \quad \gamma = \psi(0).
\end{equation*}

By definition, the residue at infinity of $f(z)$ is the coefficient of $1/z$, multiplied by $(-1)$, of the asymptotic expansion of the function $f(z)$ at infinity. It coincides with the residue
\begin{equation*}
    -\Lim{w}{0} \inv{w^2} f\br{\inv{w}}.
\end{equation*}

Before performing the computation, let us write the asymptotic expansion of the digamma function
\begin{equation*}
    \psi(z) \underset{z\to\infty}{\sim} \ln z - \inv{2z} - \sum_{k=1}^\infty \frac{B_{2k}}{2k z^{2k}},
\end{equation*}
where $B_{n}$ is the $n$-th Bernoulli number. For even values, they admit an integral representation 
\begin{equation*}
    B_{2n} = 4n (-1)^{n+1} \int_0^\infty \dd t\frac{t^{2n-1}}{e^{2\pi t}-1}.
\end{equation*}

The asymptotic expansion of $f(z)$ involves the resummation of nested series. So, we will break its computation into manageable pieces. First, we compute first the asymptotic behavior of
\begin{equation*}
    \begin{aligned}
        \psi(z) -\psi(z-2Q) &= -\ln \frac{z-2Q}{z} + \half \br{\inv{z}- \inv{z-2Q}} - \sum_{k=1}^\infty  \frac{B_{2k}}{2k }\br{\inv{z^{2k}}-\inv{(z-2Q)^{2k}}}\\
        &=- \ln \br{1 - \frac{2Q}{z}} + \inv{2z} \sbr{1- \br{1 - \frac{2Q}{z}}^{-1}}- \sum_{k=1}^\infty  \frac{B_{2k}}{2k z^{2k}}\sbr{1- \br{1 - \frac{2Q}{z}}^{-2k}}\\
        &= \sum_{n=1}^\infty \frac{1}{n} \br{ \frac{2Q}{z}}^n - \inv{2z} \sum_{n=1}^\infty \br{ \frac{2Q}{z}}^n + \sum_{k=1}^\infty  \frac{B_{2k}}{2k z^{2k}} \sum_{n=1}^\infty  \frac{\ega{2k + n}}{n! \ega{2k}}  \br{ \frac{2Q}{z}}^n,
    \end{aligned}
\end{equation*}
where we have used
\begin{equation*}
    \begin{aligned}
        \ln(1 - z) &= -\sum_{n=1}^\infty\frac{z^n}{n}, \\
        (1+z)^{-\alpha} &= \sum_{n=0}^\infty \frac{(-\alpha)(-\alpha-1)\ldots (-\alpha - n +1)}{n!}z^n = \sum_{n=0}^\infty(-1)^n \frac{\ega{\alpha + n}}{n! \ega{\alpha}} z^n.
    \end{aligned}
\end{equation*}

In multiplying the difference of digamma functions to the exponential, we must expand the exponential in power series too and match all powers to extract the coefficient of $1/z$, which gives us the residue. Let us evaluate the residue of the individual terms of $f(z)$:

\begin{align*}
    \sum_{m=0}^\infty \frac{(-1)^m}{m!} \br{i\frac{\pi p}{2Q}z^2}^{m} \sum_{n=1}^\infty \frac{1}{n} \br{ \frac{2Q}{z}}^n &\follows   \sum_{m=0}^\infty \frac{(-1)^m}{m!} \br{i\frac{\pi p}{2Q}z^2}^{m} \frac{1}{2m+1} \br{ \frac{2Q}{z}}^{2m+1} \\
    &= \frac{2Q}{z} \sum_{m=0}^\infty \frac{(-1)^m}{m!} \frac{(i\pi p 2Q)^{m}}{2m+1}\\
    &= \inv{2z}\sqrt{\frac{2Q}{i p}}  \operatorname{erf}(\sqrt{i \pi p 2 Q}).
\end{align*}
\begin{align*}
    -\sum_{m=0}^\infty \frac{(-1)^m}{m!} \br{i\frac{\pi p}{2Q}z^2}^{m}\inv{2z} \sum_{n=1}^\infty \br{ \frac{2Q}{z}}^n &\follows- \inv{2z}\sum_{m=1}^\infty \frac{(-1)^m}{m!} \br{i\frac{\pi p}{2Q}z^2}^{m}\br{ \frac{2Q}{z}}^{2m}\\
    &= -\inv{2z} \sum_{m=1}^\infty \frac{(-1)^m}{m!} (i \pi p 2Q)^{m}\\
    &= -\inv{2z} \br{e^{i\pi p 2 Q}-1} \underset{\forall p \in \integers}{=}0  .
\end{align*}
\begin{align*}
    \sum_{m=0}^\infty \frac{(-1)^m}{m!} \br{i\frac{\pi p}{2Q}z^2}^{m} &\sum_{k=1}^\infty  \frac{B_{2k}}{2k z^{2k}}  \sum_{n=1}^\infty \frac{\ega{2k + n}}{n! \ega{2k}}  \br{ \frac{2Q}{z}}^n  \follows\\
    &\follows \sum_{m=k}^\infty \frac{(-1)^m}{m!} \br{i\frac{\pi p}{2Q}z^2}^{m} \sum_{k=1}^{\infty}\frac{B_{2k}}{(2k)! z^{2k}}\frac{\ega{2m+1}}{(2m-2k+1)! }\br{ \frac{2Q}{z}}^{2m-2k+1}\\
    &= \sum_{m=1}^\infty \frac{(-1)^m}{m!} \br{i\frac{\pi p}{2Q}z^2}^{m}\sum_{k=1}^{m}\frac{B_{2k}}{(2k)! z^{2k}}
    \frac{\ega{2m+1}}{(2m-2k+1)! }\br{ \frac{2Q}{z}}^{2m-2k+1}.
    \intertext{In the last equation, we used the fact that we can either sum $m\geq k$ or $k\leq m$. The two yield the same result since the function is analytic. Expressing the even Bernoulli numbers as integrals we obtain:}
    &= -2\inv{z}\int_0^\infty \dd t \inv{e^{2\pi t} -1}\sum_{m=1}^\infty \frac{(-1)^m}{m!} \br{i\pi p 2 Q}^{m}\sum_{k=1}^{m}
    \frac{(2m)!}{(2m-2k+1)! (2k-1)!}(-1)^k \br{ \frac{t}{2Q}}^{2k-1} \\
    &=  i \inv{z}\int_0^\infty \dd t \inv{e^{2\pi t} -1}\sum_{m=0}^\infty \frac{(-1)^m}{m!} \br{i\pi p 2 Q}^{m} \sbr{\br{i\frac{t}{2Q}-1}^{2m}-\br{i\frac{t}{2Q}+1}^{2m}}\\
    &= i \inv{z}\int_0^\infty \dd t \inv{e^{2\pi t} -1} \sbr{e^{-i\frac{\pi p}{2Q}(2Q-i t)^2}-e^{-i\frac{\pi p}{2Q}(2Q+i t)^2}}\\
    &\underset{p \in \integers}{=} \inv{z}\int_0^\infty \dd t \, e^{-i\pi t} \frac{e^{-i \frac{\pi p}{2Q} t^2}}{\sinh(i \pi t)} \sinh(i 2\pi p t).
\end{align*}
This latter integral can be easily solved in the case $p=1$. Indeed consider the integral
\begin{equation}
    G_{\beta}(\alpha) = \int_0^\infty \dd t\, e^{-i \pi t} \frac{e^{-i \alpha \pi t^2}}{\sinh(i\pi t)} \sinh(i 2 \pi \beta t) ,
\end{equation}
for $\alpha \in \reals$ and $\beta\in \naturals$. For the specific value $\beta = 1$,
\begin{align*}
    G_{1}(\alpha) &= \int_0^\infty \dd t\, e^{-i \alpha \pi t^2} e^{-i\pi t} \cosh(i\pi t)\\
    &= \inv{2\sqrt{i\alpha}}\br{ e^{i \frac{\pi}{\alpha}}\br{1-\operatorname{erf}\br{{\sqrt{i\frac{\pi}{\alpha}}}}}+1}
\end{align*}
In particular, for the value $\alpha = 1/(2Q)$, which is the case $p=1$ for $Q$ integer:
\begin{equation}
    G_{1}(1/(2Q)) =  \sqrt{\frac{2Q}{i}} - \half \sqrt{\frac{2Q}{i}}\operatorname{erf}\br{\sqrt{i 2 Q}}.
\end{equation}
The latter term of the equation cancels exactly with the residue coming from the logarithmic term, when $p=1$, for any integer $Q$, yielding Eq.~\eqref{eq_phase_0_q_even}:
\begin{equation*}
    \sum_{M=0}^{2Q-1} \inv{\sqrt{2Q}} e^{-i 2\pi \frac{M^2}{4Q}} = e^{-i \pi/4}.
\end{equation*}

Somehow more difficult is the evaluation for $p\neq 1$ and $p$ odd and coprime with $2Q$. 

\section{Free Qubit Hamiltonian}\label{a_adimensional_free_hamiltonian}

For the free one-particle Hamiltonian considered the main text, the main technical difficulty in its study is the momenta spacing: physical momenta are spaced as $\momentumlattice{L}$, on a circle of length $L$. On a digital quantum simulation with $n$ qubits, it is natural to combine them as in Sec.~\ref{s_rev_algo} to construct $N = 2^n$ states, which conveniently assume the label of position. However, if we treat these states as an abstract basis for any finite sequence $\cbr{f_J}$, $J = \cbr{-N/2, \ldots, N/2-1}$, we can introduce $N$ adimensional momenta $k_M$, spaced $2\pi/N$ and spanning the range
\begin{equation}
    k \in \cbr{- \pi , -\pi + \frac{2\pi}{N}, \ldots, \pi -\frac{2\pi}{N}} \defi \frac{\cbr{-\pi, \pi}}{N}. 
\end{equation}
In the limit $N \to\infty$, this setup approaches the geometry of an infinite lattice of unit spacing. 

If one would like to connect these adimensional momenta with the physical ones $\kappa$, as long as both $N$ and $L$ are finite, they are related by 
\begin{equation}
    \kappa = \frac{N}{L} k = \inv{\ell} k.
\end{equation}
As such, we propose a "free qubit Hamiltonian" $K_N$ which utilizes, instead of the $\kappa$'s, the $k$'s which have been previously introduced:
\begin{equation}
    K_N = \sum_{k\in \cbr{-\pi, \pi}/N} \frac{k^2}{2}\pro{k}.
\end{equation}
First of all, this kinetic term is adimensional and bounded, contrary to the kinetic energy on the continuous circle, since the $\kappa$'s are discrete. This property makes its propagator amenable to straightforward numerical computations and easy analytical manipulations. 

For completeness, let $\cbr{f_J}$ be a sequence of $N$ complex numbers. From it, one can construct the many-body wave function
\begin{equation}
    \ket{f} = \sum_{J = -N/2}^{N/2-1} f^J \ket{J}.
\end{equation}
where the $J$'s are defined as in Sec.~\ref{s_rev_algo}. The basis of $k$'s is complete and related to that of $J$'s through the quantum Fourier Transform
\begin{equation}
    \ket{k} = \sum_{J= -N/2}^{N/2-1} \frac{e^{i k J}}{\sqrt{N}} \ket{J}, \quad \ket{J} = \sum_{k\in \cbr{-\pi, \pi}/N} \frac{e^{-i k J}}{\sqrt{N}} \ket{k}.
\end{equation}
This map, being unitary, induces a discrete Fourier Transform on the $f_J$ into some $F_k$:
\begin{equation}
    f_J =\sum_{k\in \cbr{-\pi, \pi}/N} \frac{e^{i k J}}{\sqrt{N}} F_k , \quad F_k = \sum_{J= -N/2}^{N/2-1}  \frac{e^{-i k J}}{\sqrt{N}} f_J.
\end{equation}
Studying the ``free qubit'' propagator
\begin{equation}\label{eq_free_qubit_propagator}
    D_J^{J'}(\tau) = \braket{J'|e^{-i K_N \tau}|J}= \sum_{k\in \cbr{-\pi, \pi}/N}\inv{N} \exp\cbr{- i k^2 \frac{\tau}{2} + i k \br{J'-J}},
\end{equation}
symmetric in $(J'-J)\to -(J'-J)$, is a necessary step to study the time evolution of states $\ket{f}$ under Hamiltonians of the form
\begin{equation}
    H_N = K_N + V_N,
\end{equation}
where $V_N$ are potential terms diagonal in the basis $\ket{J}$. Naturally, one can consider more general potentials, involving non-local interactions, i.e. terms in the Hamiltonians which have nonzero matrix entries corresponding to the projectors $\ket{J'}\bra{J}$. The time evolution follows closely Eq.~\eqref{eq_time_evo_circuit}. Since $K_N$ is a quadratic operator, $e^{-i K_N \tau}$ can be constructed with $O(n^2)$ perfect gates, while the potential depends on the particular function chosen \cite{MST2024}.

As such, in the following, we will consider the dynamics of a ``localised'' qubit state $\ket{J}$ evolved under $K_N$. For any finite $N$, the computation can be performed numerically, but an analytic form can be obtained in the limit of large $N$, for which the sum in Eq.~\eqref{eq_free_qubit_propagator} can be approximated by an integral, which in turn defines an analytic continuation to $(J'-J)$ outside the integers:
\begin{equation}\label{eq_qubit_propagator_large_N}
\begin{aligned}
    \mathcal{D}_J^{J'}(\tau) &= \braket{J'|e^{-i K_\infty \tau}|J}= \inv{2\pi} \int_{-\pi}^\pi \dd k \, \exp\cbr{- i k^2 \frac{\tau}{2} + i k \br{J'-J}} \\
    &= \inv{2\pi} \frac{\sqrt{\pi}}{2}\frac{e^{i\frac{(J'-J)^2}{2\tau}}}{\sqrt{i \tau/2}} \sbr{\operatorname{erf} \br{\frac{i(J'-J)}{2\sqrt{i \tau/2}}+\pi \sqrt{i \tau/2} } - \operatorname{erf}\br{\frac{i(J'-J)}{2\sqrt{i \tau/2}} -\pi \sqrt{i \tau/2}}} \\
    &= \inv{2\pi} \frac{\sqrt{\pi}}{2}\frac{e^{i\frac{(J'-J)^2}{2\tau}}}{\sqrt{i \tau/2}} \sbr{\operatorname{erf} \br{\pi \sqrt{i \tau/2} +\frac{i(J'-J)}{2\sqrt{i \tau/2}}} + \operatorname{erf}\br{\pi \sqrt{i \tau/2}-\frac{i(J'-J)}{2\sqrt{i \tau/2}}}}.
    \end{aligned}
\end{equation}
Here $\operatorname{erf}$ denotes the error function 
\begin{equation*}
    \operatorname{erf} (z) = \frac{2}{\sqrt{\pi}} \int_0^z \dd t\, e^{-t^2} = \frac{2}{\sqrt{\pi}} \sum_{n=0}^\infty \frac{(-1)^n z^{2n+1}}{n! (2n+1)}.
\end{equation*}
satisfying the following properties
\begin{equation*}
\begin{aligned}
    \int_0^a \dd k\, e^{-\alpha k^2 + \beta k} &= \frac{e^{\beta^2/(4\alpha)}}{2\sqrt{\alpha/\pi}} \sbr{\operatorname{erf} \br{a \sqrt{\alpha}-\frac{\beta}{2\sqrt{\alpha}} } - \operatorname{erf}\br{-\frac{\beta}{2\sqrt{\alpha}} }}\\
    &= -\frac{e^{\beta^2/(4\alpha)}}{2\sqrt{\alpha/\pi}} \sbr{\operatorname{erf} \br{\frac{\beta}{2\sqrt{\alpha}} -a\sqrt{\alpha} }-\operatorname{erf}\br{\frac{\beta}{2\sqrt{\alpha}} }},\\
    \int_{-a}^a \dd k\, e^{-\alpha k^2 + \beta k} &= \frac{e^{\beta^2/(4\alpha)}}{2\sqrt{\alpha/\pi}} \sbr{\operatorname{erf} \br{\frac{\beta}{2\sqrt{\alpha}}+a \sqrt{\alpha} } - \operatorname{erf}\br{\frac{\beta}{2\sqrt{\alpha}} -a \sqrt{\alpha}}}.
    \end{aligned}
\end{equation*}
We immediately recognize
\begin{equation}
    S_{J}^{J'}(\tau) = \frac{(J'-J)^2}{2\tau}
\end{equation}
to be the equal in form to the action (when $\hbar =1$) of a free particle of unit mass moving between $J$ and $J'$ in time $\tau$. In our opinion, it is this appealing connection that  makes this free qubit Hamiltonian worth studying. 

We will study separately the small-time behavior of the propagator, useful for example in the path integral formulation of the time evolution
\begin{equation}
    \braket{J'|e^{-i H_N \tau }|f} = \intmpinf{J_\mathcal{N}} \cdots\intmpinf{J_2} \intmpinf{J_1}\, D_{J_\mathcal{N}}^{J'}\br{\frac{\tau}{\mathcal{N}}} e^{-i V_{J_\mathcal{N}} \frac{\tau}{\mathcal{N}}} \cdots D_{J_2}^{J_3}\br{\frac{\tau}{\mathcal{N}}} e^{-i V_{J_2} \frac{\tau}{\mathcal{N}}}D_{J_1}^{J_2}\br{\frac{\tau}{\mathcal{N}}} e^{-i V_{J_1} \frac{\tau}{\mathcal{N}}}f^{J_1},
\end{equation}
where $\tau/\mathcal{N}\ll 1$,
and the large-time behavior, which turns out to be uniformly distributed, but if compared to Eq.~\eqref{e_propa_line_limit} there is a cut-off in the maximum velocity of propagation, i.e. $\abs{(J'-J)} \leq \pi \tau$.

\subsection{Small-Time Behavior}

For small times, considering the two quantities $\frac{(J'-J)}{2\sqrt{\tau/2}}$ and $\pi \sqrt{ \tau/2}$, we can treat the second as the leading term and the first as a perturbation. Since in the propagator we have a subtraction, only odd terms in the series appear. Also recall that the first derivative of the error function is the Gaussian, therefore
\begin{equation}
    \mathcal{D}_J^{J'}(\tau) = \sum_{n=0}^\infty H_{2n}\br{\frac{i(J'-J)}{2\sqrt{i \tau/2}}} \frac{\pi^{2n} (i \tau /2)^n}{(2n+1)!},
\end{equation}
where the functions $H_{n}(x)$ are the Hermite polynomials
\begin{equation*}
    H_{n}(x) = e^{x^2} \dv_x^n e^{-x^2} =n! \sum_{m=0}^{\floor{n/2}} \frac{(-1)^m}{m! (n-2m)!} (2x)^{n-2m}
\end{equation*}
and $\floor{n}$ is the truncation of the number $n$ from below.

The series for $\mathcal{D}$ is not yet expressed in terms of powers of $\tau$, for a very simple reason: that the Hermite polynomials depend on $\tau$ as well. We then have to consider, term-by-term the individual contributions to a fixed order in $\tau$, and sum them. First of all notice that the Hermite polynomial $H_n(x)$ will always be of the form
\begin{equation*}
    H_n(x)\sim a_0^{(n)} x^{n} + a_{1}^{(n)} x^{n-2} + a_{2}^{(n)} x^{n-4} + \ldots.
\end{equation*}
Moreover, the term $a_{n-2}$ appears only for $n\geq 2$, the term $a_{n-4}$ for $n\geq 4$ and so on. Also the coefficient $a_{n-2m}$ is given by
\begin{equation*}
    a_{m}^{(n)} = n!\frac{(-1)^m}{m! (n-2m)!}.
\end{equation*}
In other words, $a_0$ will appear in any $H_{n}$, $a_1$ will appear only starting from $H_{2}$, $a_2$ from $H_{4}$ and so on. We may then write $\mathcal{D}$ as
\begin{equation}
\begin{aligned}
    \mathcal{D}_J^{J'}(\tau) &=  \sum_{m=0}^{\infty}\sum_{n=m}^{\infty} a_m^{(2n)} \br{2\frac{i(J'-J)}{2\sqrt{i\tau/2}}}^{2(n-m)} \frac{\pi^{2n}(i\tau/2)^n}{(2n+1)!}\\
    &= \sum_{m=0}^{\infty} \frac{(-i\pi^2 \tau/2)^m}{m!} \sum_{n=0}^{\infty} \frac{1}{2n+1 + 2m} \frac{(-1)^n \br{\pi(J'-J)}^{2n}}{(2n)!}\\
    &= \sum_{m=0}^{\infty} \frac{(-i\pi^2 \tau/2)^m}{m!}\mathcal{I}_{2m} (\pi (J'-J)) \\
    &= \sum_{m=0}^{\infty} \frac{(-1)^m (\pi^2 \tau/2)^{2m}}{(2m)!}\mathcal{I}_{4m} (\pi (J'-J))-i\sum_{m=0}^{\infty} \frac{(-1)^{m} (\pi^2 \tau/2)^{2m+1}}{(2m+1)!}\mathcal{I}_{2(2m + 1)} (\pi (J'-J)).
\end{aligned}
\end{equation}
In the last line we separated the real and imaginary part. The propagator is then expressed in terms of some functions $\mathcal{I}_{m}$, which in turn are related to some other functions $\mathcal{J}_{m}$: 
\begin{equation}
\begin{aligned}
   \mathcal{I}_m (x) &= \sum_{n=0}^{\infty} \frac{1}{2n+1 +m} \frac{(-1)^n x^{2n}}{(2n)!}, \quad \mathcal{I}_0 (x) = \frac{\sin (x)}{x},\\
   \mathcal{J}_m (x) &= \sum_{n=0}^{\infty} \frac{1}{2(n+1) +m} \frac{(-1)^n x^{2n}}{(2n+1)!}, \quad \mathcal{J}_0 (x) = \frac{1-\cos(x)}{x^2}.
   \end{aligned}
\end{equation}
We can get a recursive formula for $\mathcal{I}_m$ and $\mathcal{J}_m$. Because of 
\begin{equation*}
\begin{aligned}
   \mathcal{I}_m (x) &= \sum_{n=0}^{\infty}\br{ \frac{1}{2n+1 +m}-\frac{1}{2n+1}} \frac{(-1)^n x^{2n}}{(2n)!} + \mathcal{I}_0 (x)\\
   & =  -m \sum_{n=0}^{\infty}\frac{1}{2n+1 +m}\frac{(-1)^n x^{2n}}{(2n+1)!} + \mathcal{I}_0 (x)\\
   &= -m \mathcal{J}_{m-1} (x) + \mathcal{I}_0 (x),\\
   \mathcal{J}_m (x) &= \sum_{n=0}^{\infty} \br{\frac{1}{2(n+1) +m} -\frac{1}{2(n+1)}} \frac{(-1)^n x^{2n}}{(2n+1)!}+ \mathcal{J}_0 (x) \\
   &= -m\sum_{n=0}^{\infty} \frac{1}{2(n+1) +m} \frac{(-1)^n x^{2n}}{(2n+2)!}+ \mathcal{J}_0 (x) \\
   &= \frac{m}{x^2}\sum_{n=0}^{\infty} \frac{1}{2(n+1) +1 +(m-1)} \frac{(-1)^{n+1} x^{2(n+1)}}{(2(n+1))!}+ \mathcal{J}_0 (x) \\
   &= \frac{m}{x^2}\sum_{n=1}^{\infty} \frac{1}{2n +1 +(m-1)} \frac{(-1)^{n} x^{2n}}{(2n)!}+ \mathcal{J}_0 (x)\\
   &= \frac{m}{x^2}\br{\mathcal{I}_{m-1}-\inv{m}}+ \mathcal{J}_0 (x),
   \end{aligned}
\end{equation*}
we can write
\begin{equation}
   \mathcal{I}_{2m} (x) = \mathcal{I}_0 (x) - 2m \frac{1-\cos(x)}{x^2} - \frac{2m(2m-1)}{x^2}\br{\mathcal{I}_{2(m-1)} (x) - \frac{1}{2m-1}}, \quad \mathcal{I}_0 (x) = \frac{\sin(x)}{x}.
\end{equation}
For completeness, we list the first three entries:
\begin{align*}
    \mathcal{I}_0 (x) &= \frac{\sin(x)}{x}\\
    \mathcal{I}_2 (x) &= \mathcal{I}_0 (x) - 2 \mathcal{J}_0 (x) + \frac{2}{x^2} \br{1-\mathcal{I}_0 (x)} \\
    &= \frac{\sin (x)}{x} +\frac{2 \cos (x)}{x^2} -\frac{2 \sin (x)}{x^3}\\
    \mathcal{I}_4 (x) &= \mathcal{I}_0 (x) - 4 \mathcal{J}_0 (x) +\frac{12}{x^2} \br{1-\mathcal{I}_0 (x)} -\frac{24}{x^2}\br{\half - \mathcal{J}_0(x) } + \frac{24}{x^4} \br{\mathcal{I}_0-1+\frac{x^2}{6}}\\
    &= \frac{\sin (x)}{x} +\frac{4 \cos (x)}{x^2} -\frac{12 \sin (x)}{x^3} -\frac{24 \cos (x)}{x^4} +  \frac{24 \sin (x)}{x^5}.
\end{align*}

It is easy to evaluate these functions when $x= \pi (J'-J)$, for $(J'-J)\in \integers$. For $J'=J$, we only look at the leading contributions of the series defining $\mathcal{I}_{2m}$, while for $J'\neq J$ the recursive relation simplifies to
\begin{equation*}
    \mathcal{I}_{2m}(x)|_{x = \pi (J'-J)} = \sbr{2m \frac{e^{ix}}{x^2} - \frac{2m(2m-1)}{x^2} \mathcal{I}_{2m-2} (x)}_{x = \pi (J'-J)}.
\end{equation*}
Solving the recursive equation yields, 
\begin{equation}
    \mathcal{I}_{2m}(\pi (J'-J)) = \frac{1}{2m+1} \krdel{J'}{J} + \br{1-\krdel{J'}{J}} \frac{\cos(\pi (J'-J))}{\sbr{\pi (J'-J)}^{2}}  \sum_{n=0}^{m-1} \frac{(-1)^{m-1-n} (2m)! }{(2n+1)! \sbr{\pi (J'-J)}^{2(m-1-n)}},
\end{equation}
where $\cos(\pi (J'-J))$ assumes identically the values $(-1)^{(J'-J)}$. The second term of the equation only appears for $m\geq 2$. As such, the first three of these read: 
\begin{equation}
    \begin{aligned}
        \mathcal{I}_0 (\pi (J'-J)) &= \krdel{J'}{J},\\
        \mathcal{I}_2 (\pi (J'-J)) &= \frac{1}{3}\krdel{J'}{J} + \br{1-\krdel{J'}{J}} \cos(\pi (J'-J))\frac{2}{\pi^2 (J'-J)^2},\\
        \mathcal{I}_4 (\pi (J'-J)) &= \frac{1}{5}\krdel{J'}{J} + \br{1-\krdel{J'}{J}} \cos(\pi (J'-J))\br{\frac{4}{\pi^2 (J'-J)^2}-\frac{24}{\pi^4 (J'-J)^4}}.
    \end{aligned}
\end{equation}

\begin{figure}
    \centering
    \includegraphics[scale =0.75]{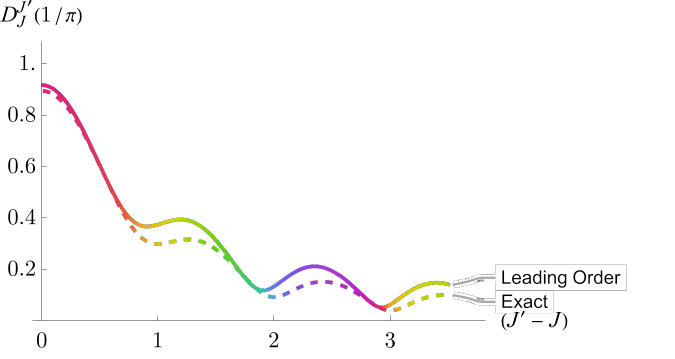}
    \caption{Plot of the modulus and phase of the free qubit propagator as function of $(J'-J)$ Eq.~\eqref{eq_qubit_propagator_large_N} (dashed line) and of the leading approximation Eq.~\eqref{eq_propagator_leading_approximation} (solid line) at time $\tau = 1/\pi$. The time is chosen sufficiently large that the differences are noticeable.}
    \label{f_propagatorsmallt}
\end{figure}

Summarizing, the short-time behavior of the propagator is:
\begin{equation}\label{eq_propagator_leading_approximation}
    \mathcal{D}_J^{J'}(\tau) = \br{\mathcal{I}_0 (\pi (J'-J)) - \frac{\pi^4}{8} \mathcal{I}_4 (\pi (J'-J)) \tau^2} - i\frac{\pi^2}{2} \mathcal{I}_2 (\pi (J'-J)) \tau.
\end{equation}
Keeping in mind that only $(J'-J)\in \integers$ is a possible outcome of a simulation, the limiting procedure allows us to provide an analytic continuation in the reals. Numerically, deviations from this approximation are appreciable for times $\tau > 0.2$, see Fig.~\ref{f_propagatorsmallt}.

The analysis of the propagator at short times is complete if we provide expressions for the modulus and phase. Let us start from the modulus of the propagator, which is related to the transition probability. It reads
\begin{equation}
\begin{aligned}
    \abs{\mathcal{D}_J^{J'}(\tau)} &= \sbr{\sbr{\mathcal{I}_0 (\pi (J'-J))}^2 - \tau^2 \frac{\pi^4}{4}\sbr{ \mathcal{I}_0 (\pi (J'-J))\mathcal{I}_4 (\pi (J'-J))-\br{\mathcal{I}_2 (\pi (J'-J))}^2}}^{\half}.
\end{aligned}
\end{equation}
This modulus is positive for sufficiently small values of $\tau$, which is fixed by requiring that the probability of return is non-vanishing
\begin{equation}
    \norm{\mathcal{D}_J^{J}(\tau)}  =1 - \tau^2 \frac{\pi^4}{45}\geq 0 \follows 
    \tau \leq \frac{\sqrt{45}}{\pi^2} \approx 0.67868.
\end{equation}
This latter time is an upper bound for the validity of the small-time expansion. Also using the exact results for $\br{J'-J}$ to be a nonzero integer, the corresponding transition probability is 
\begin{equation}
    \norm{\mathcal{D}_J^{J'}(\tau)}  =\frac{\tau^2}{(J'-J)^4} = \frac{4}{\br{S_{J}^{J'}(\tau)}^2}
\end{equation}
As a check of consistency, we verify that probability is conserved:
\begin{equation}
\begin{aligned}
    \sum_{J' = -\infty}^\infty  \norm{\mathcal{D}_J^{J'}(\tau)}&= 1 - \tau^2 \br{\frac{\pi^4}{45}  -\sum_{j = -\infty,\, j\neq 0}^\infty \frac{1}{(J'-J)^4}} + O(\tau^4) \\
    &= 1 - \tau^2 \br{\frac{\pi^4}{45}  -2 \zeta(4)} + O(\tau^4)\\
    &\underset{\zeta(4) = \pi^4/90}{=} 1+O(\tau^4).
\end{aligned}
\end{equation}
On the other hand, due to the interplay between series and integration, we also claim a nontrivial result, i.e. that
\begin{equation}
    \int_{-\infty}^\infty \dd J'\, \norm{\mathcal{D}_{J}^{J'}(\tau)}  = 1 + O(\tau^4) \follows \int_{-\infty}^\infty \dd J'\, \sbr{ \mathcal{I}_0 (\pi (J'-J))\mathcal{I}_4 (\pi (J'-J))-\br{\mathcal{I}_2 (\pi (J'-J))}^2} = 0.
\end{equation}
Therefore, for very small $\tau$, $2\zeta(4)\tau^2$ is the probability that we do not detect any particle at the origin after time $\tau$. For example, $\tau^2/n^4$ is the probability of detecting $(J'-J) = \pm n$.

If compared to the propagator on the circle, presented in the main text, we see that small-time behavior interpolates between a Kroneker delta, paradigmatic of discrete positions, and a regime in which $(J'-J) \neq 0$ become more probable as time increases. The physical interpretation is that of providing a transient between the zero-time and infinitesimal time behavior of the propagator on the line Eq.~\eqref{e_propa_line_limit}.

Let us now consider the phase of the leading approximation Eq.~\eqref{eq_propagator_leading_approximation}. Formally we can obtain it by considering the angle defined by the imaginary and real part of the complex number, i.e.
\begin{equation*}
    \varphi := \arctan\br{\frac{-\frac{\pi^2}{2}\mathcal{I}_2 \br{\pi (J'-J)} \tau}{\mathcal{I}_0 \br{\pi (J'-J)} - \frac{\pi^4}{8}\mathcal{I}_4 \br{\pi (J'-J)}\tau^2}}.
\end{equation*}
However, this formulation is highly impenetrable and we would like to relate it with some quantity involving the free action $S = S^{J'}_{J}(\tau)$. We shall consider first the case $(J'-J)= 0$, where no dependence on position is expected.

For $(J'-J)= 0$ we use an exact resummation. Recall that $\mathcal{I}_{2m} (0) =1/\br{2m+1}$. Moreover, let us introduce the Fresnel integrals $\mathcal{S}(x)$ and $\mathcal{C}(x)$:
\begin{equation}
    \begin{aligned}
        \mathcal{C}(z) &= \int_0^z \dd t\, \cos\br{\half \pi t^2} = \sum_{n=0}^\infty \frac{(-1)^n (\pi/2)^{2n}}{(2n)!(4n+1)}z^{4n+1}, \\
        \mathcal{S}(z) &=  \int_0^z \dd t\, \sin\br{\half \pi t^2}= \sum_{n=0}^\infty \frac{(-1)^n (\pi/2)^{2n+1}}{(2n+1)!(4n+3)}z^{4n+3}.
    \end{aligned}
\end{equation}
It is easy to cast the propagator at $J'=J$ as
\begin{equation}
    \mathcal{D}^{J}_{J}(\tau) = \frac{\mathcal{C}(\sqrt{\pi\tau})}{\sqrt{\pi\tau}} - i \frac{\mathcal{S}(\sqrt{\pi\tau})}{\sqrt{\pi\tau}} = \frac{\operatorname{erf} (\pi\sqrt{i\tau/2})}{\sqrt{i 2\pi \tau}}.
\end{equation}
In turn, the last equation allows us to find the identity:
\begin{equation}
        \operatorname{erf}(\sqrt{i z}) = \sqrt{2i}\br{\mathcal{C}(\sqrt{2z/\pi})-i \mathcal{S}(\sqrt{2 z/\pi})}, \quad z\in \reals^+.
\end{equation}
On the other hand, since the action for $J'=J$ is always vanishing, we can write that at least for the transition amplitude of return:
\begin{equation}
    \mathcal{D}^{J}_{J}(\tau) = \frac{e^{i S}}{\sqrt{i 2\pi \tau}}  \operatorname{erf} (\pi\sqrt{i\tau/2}),
\end{equation}
with $S = (J'-J)/2\tau$.
Therefore the phase is 
\begin{equation}
    \varphi_{J}^{J} (\tau) = S - \arctan \frac{\mathcal{S}(\sqrt{\pi\tau})}{\mathcal{C}(\sqrt{\pi\tau})}.
\end{equation}
For small enough times, the phase goes as 
\begin{equation}
    \varphi_{J}^{J} (\tau) \sim S - \frac{\pi^2}{6}\tau,
\end{equation}
while for large values, since
\begin{equation*}
    \begin{aligned}
        \mathcal{S}(\sqrt{\pi \tau})&= \half - \frac{\cos\br{\frac{\pi ^2}{2}\tau}}{\pi \sqrt{\pi \tau}},\\
        \mathcal{C}(\sqrt{\pi \tau})&= \half + \frac{\sin\br{\frac{\pi ^2}{2}\tau}}{\pi \sqrt{\pi \tau}}
    \end{aligned}
\end{equation*}
and $\arctan 1 = \pi/4$,
\begin{equation}
    \varphi_{J}^{J} (\tau) \sim S - \frac{\pi}{4} + \frac{2\sbr{\cos\br{\frac{\pi ^2}{2}\tau}+\sin\br{\frac{\pi ^2}{2}\tau}}}{\pi \sqrt{\pi \tau}} + O(1/\tau).
\end{equation}
This phase asymptotizes, at least for $(J'-J)=0$, to that of the particle on the line Eq.~\eqref{e_propa_line_limit}.

\subsection{Large-Time Behavior}

\begin{figure}
    \centering
    \includegraphics[scale=0.75]{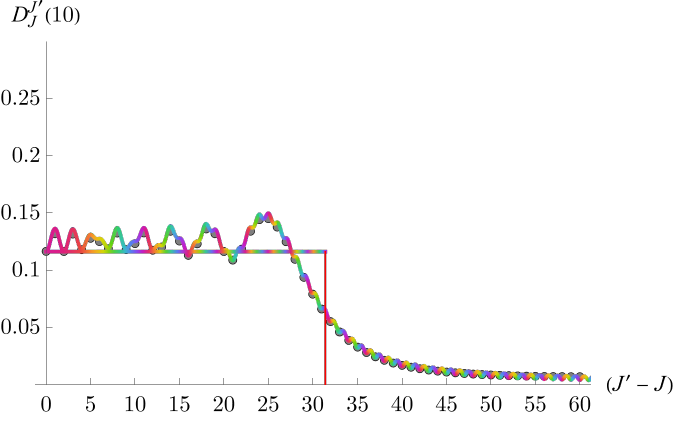}
    \caption{Plot of the modulus of the propagator, the values of the exact one for $N=120$ (grey dots) and the uniform approximation (horizontal line) Eq.~\eqref{e_propa_line_limit} for a time $\tau = 10$. The red vertical line is in correspondence of the position $\abs{J'-J} = \pi \tau$. The phase corresponds to the line colors and is well-captured by that of the Feynman propagator for sufficiently large times.}
    \label{f_propagator_discrete_continuous}
\end{figure}

In light of the latter rewriting of the propagator involving Fresnel integrals, we can rewrite the propagator by substituting $(J'-J)^2/2\tau$ with the action $S$:
\begin{equation*}
    \mathcal{D}_J^{J'}(\tau) =\frac{e^{i S}}{\sqrt{i 2\pi  \tau}} \half\sbr{\operatorname{erf} \br{\pi \sqrt{i \tau/2} + \sqrt{iS}} +\operatorname{erf}\br{\pi \sqrt{i \tau/2}-\sqrt{iS}}}.
\end{equation*}
We can find two expressions in terms of the real and imaginary parts. Since the argument of the second error function depends on the difference between $\sqrt{\pi^2 \tau/2}$ and $\sqrt{S}$, we have that whenever the former is greater than the latter, i.e. $\abs{J'-J}\leq \pi \tau$:
\begin{equation}
\begin{aligned}
    \mathcal{D}_J^{J'}(\tau) &= \frac{e^{i S}}{2\sqrt{\pi  \tau}} \Bigg\{ \sbr{ \mathcal{C}\br{\sqrt{\pi \tau + \frac{2S}{\pi} + 2\sqrt{2\tau S}}} + \mathcal{C}\br{\sqrt{\pi \tau + \frac{2S}{\pi} - 2\sqrt{2\tau S}}}}\\
    &\qquad -i \sbr{\mathcal{S}\br{\sqrt{\pi \tau + \frac{2S}{\pi} + 2\sqrt{2\tau S}}} + \mathcal{S}\br{\sqrt{\pi \tau + \frac{2S}{\pi} - 2\sqrt{2\tau S}}}} \Bigg\}, \quad \abs{J'-J}<\pi \tau.
    \end{aligned}
\end{equation}
On the other hand, if $\abs{J'-J} \geq \pi \tau$:
\begin{equation}
\begin{aligned}
    \mathcal{D}_J^{J'}(\tau) &= \frac{e^{i S}}{2\sqrt{\pi  \tau}} \Bigg\{ \sbr{ \mathcal{C}\br{\sqrt{\pi \tau + \frac{2S}{\pi} + 2\sqrt{2\tau S}}} -\mathcal{C}\br{\sqrt{\pi \tau + \frac{2S}{\pi} - 2\sqrt{2\tau S}}}}\\
    &\qquad -i \sbr{\mathcal{S}\br{\sqrt{\pi \tau + \frac{2S}{\pi} + 2\sqrt{2\tau S}}} - \mathcal{S}\br{\sqrt{\pi \tau + \frac{2S}{\pi} - 2\sqrt{2\tau S}}}} \Bigg\}, \quad \abs{J'-J}>\pi \tau.
    \end{aligned}
\end{equation}
This manipulation allows us to find the phase $\varphi$ for generic values of $S$ and $\tau$:
\begin{equation}
    \varphi^{J'}_{J} (\tau) =\begin{cases}
        S - \arctan\frac{\mathcal{S}\br{\sqrt{\pi \tau + \frac{2S}{\pi} + 2\sqrt{\tau S}}} + \mathcal{S}\br{\sqrt{\pi \tau + \frac{2S}{\pi} - 2\sqrt{2\tau S}}}}{\mathcal{C}\br{\sqrt{\pi \tau + \frac{2S}{\pi} + 2\sqrt{2\tau S}}} + \mathcal{C}\br{\sqrt{\pi \tau + \frac{2S}{\pi} - 2\sqrt{2\tau S}}}}, & \abs{J'-J}<\pi \tau,\\
        S - \arctan\frac{\mathcal{S}\br{\sqrt{\pi \tau + \frac{2S}{\pi} + 2\sqrt{\tau S}}} - \mathcal{S}\br{\sqrt{\pi \tau + \frac{2S}{\pi} - 2\sqrt{2\tau S}}}}{\mathcal{C}\br{\sqrt{\pi \tau + \frac{2S}{\pi} + 2\sqrt{2\tau S}}} -\mathcal{C}\br{\sqrt{\pi \tau + \frac{2S}{\pi} - 2\sqrt{2\tau S}}}}, & \abs{J'-J}>\pi \tau.
    \end{cases} 
\end{equation}
The splitting suggests that the propagation has actually two competing modes: $\abs{J'-J} <\pi \tau$ and $\abs{J'-J} >\pi \tau$. For $\abs{J'-J} <\pi \tau$, as time goes on, the propagator will tend more and more to a uniform distribution, while for $\abs{J'-J} >\pi \tau$, the propagator will be more and more suppressed. As such, the picture we can glimpse at large times is that of a uniform propagator that is non-vanishing only for $\abs{J'-J} <\pi \tau$. These qualitative features are visualized in Fig.~\ref{f_propagator_discrete_continuous}.

Asymptotic analysis of this propagator suggests that for large times
\begin{equation}
    \mathcal{D}_J^{J'}(\tau) = \frac{e^{i S_{J}^{J'}(\tau)}}{\sqrt{i 2\pi \tau}}\br{1+O(1/\sqrt{\tau})}, \quad \abs{J'-J}\leq \pi \tau,
\end{equation}
and zero otherwise. Propagation happens within a light-cone $\abs{J'-J}\leq \pi \tau$ dictated by the maximum velocity allowed $k = \pm \pi$. It would be interesting to study further this propagator, in relation to the well-known fact that also a relativistic particle can propagate within a light-cone.

\end{document}